\documentclass[twocolumn,numberedappendix,twocolappendix,iop]{openjournal}

\usepackage{amsmath}
\usepackage{lineno}
\usepackage{xcolor}
\usepackage[T1]{fontenc}
\usepackage[colorlinks,linkcolor=blue,citecolor=blue,urlcolor=blue]{hyperref}
\usepackage{graphicx}	
\usepackage{color,soul}
\usepackage{natbib}
\defcitealias{Koposov2024_EDR_VAC}{K24}

\newcommand{\kms}{\,km\,$^{-1}$}

\begin{document}
\title{DESI Data Release 1: Stellar Catalogue}

\def\andname{}

\author{
\vspace{-1.1cm}
Sergey~E.~Koposov\altaffilmark{1,2},
Ting~S.~Li\altaffilmark{3},
C.~Allende~Prieto\altaffilmark{4,5},
G.~E.~Medina\altaffilmark{3},
N.~Sandford\altaffilmark{3},
D.~Aguado\altaffilmark{5},
L.~{Beraldo e Silva}\altaffilmark{6,7,8},
A.~Bystr\"om\altaffilmark{1},
A.~P.~Cooper\altaffilmark{9},
Arjun~Dey\altaffilmark{10},
C.~S.~Frenk\altaffilmark{11},
N.~Kizhuprakkat\altaffilmark{9},
S.~Li\altaffilmark{12},
J.~Najita\altaffilmark{10},
A.~H.~Riley\altaffilmark{11},
D.~R.~Silva\altaffilmark{13},
G.~Thomas\altaffilmark{5},
M.~Valluri\altaffilmark{6,14},
J.~Aguilar\altaffilmark{15},
S.~Ahlen\altaffilmark{16},
D.~Bianchi\altaffilmark{17,18},
D.~Brooks\altaffilmark{19},
T.~Claybaugh\altaffilmark{15},
S.~Cole\altaffilmark{11},
A.~Cuceu\altaffilmark{15},
A.~de la Macorra\altaffilmark{20},
J.~Della~Costa\altaffilmark{21,10},
Biprateep~Dey\altaffilmark{3,22},
P.~Doel\altaffilmark{19},
J.~Edelstein\altaffilmark{23,24},
A.~Font-Ribera\altaffilmark{25},
J.~E.~Forero-Romero\altaffilmark{26,27},
E.~Gaztañaga\altaffilmark{28,29,30},
S.~Gontcho A Gontcho\altaffilmark{15},
G.~Gutierrez\altaffilmark{31},
J.~Guy\altaffilmark{15},
K.~Honscheid\altaffilmark{32,33,34},
J.~Jimenez\altaffilmark{25},
R.~Kehoe\altaffilmark{35},
D.~Kirkby\altaffilmark{36},
T.~Kisner\altaffilmark{15},
A.~Kremin\altaffilmark{15},
O.~Lahav\altaffilmark{19},
M.~Landriau\altaffilmark{15},
L.~Le~Guillou\altaffilmark{37},
A.~Leauthaud\altaffilmark{38,39},
M.~E.~Levi\altaffilmark{15},
M.~Manera\altaffilmark{40,25},
A.~Meisner\altaffilmark{10},
R.~Miquel\altaffilmark{41,25},
J.~Moustakas\altaffilmark{42},
S.~Nadathur\altaffilmark{29},
N.~Palanque-Delabrouille\altaffilmark{43,15},
W.~J.~Percival\altaffilmark{44,45,46},
F.~Prada\altaffilmark{47},
I.~P\'erez-R\`afols\altaffilmark{48},
G.~Rossi\altaffilmark{49},
E.~Sanchez\altaffilmark{50},
E.~F.~Schlafly\altaffilmark{51},
D.~Schlegel\altaffilmark{15},
H.~Seo\altaffilmark{52},
R.~Sharples\altaffilmark{53,11},
J.~Silber\altaffilmark{15},
D.~Sprayberry\altaffilmark{10},
G.~Tarl\'{e}\altaffilmark{14},
B.~A.~Weaver\altaffilmark{10},
R.~Zhou\altaffilmark{15},
and H.~Zou\altaffilmark{54}
 \\
}

\altaffiltext{1}{Institute for Astronomy, University of Edinburgh, Royal Observatory, Blackford Hill, Edinburgh EH9 3HJ, UK}
\altaffiltext{2}{Institute of Astronomy, University of Cambridge, Madingley Road, Cambridge CB3 0HA, UK}
\altaffiltext{3}{Department of Astronomy \& Astrophysics, University of Toronto, Toronto, ON M5S 3H4, Canada}
\altaffiltext{4}{Departamento de Astrof\'{\i}sica, Universidad de La Laguna (ULL), E-38206, La Laguna, Tenerife, Spain}
\altaffiltext{5}{Instituto de Astrof\'{\i}sica de Canarias, C/ V\'{\i}a L\'{a}ctea, s/n, E-38205 La Laguna, Tenerife, Spain}
\altaffiltext{6}{Department of Astronomy, University of Michigan, Ann Arbor, MI 48109, USA}
\altaffiltext{7}{Steward Observatory, University of Arizona, 933 N. Cherry Avenue, Tucson, AZ 85721, USA}
\altaffiltext{8}{Observatório Nacional, Rio de Janeiro - RJ, 20921-400, Brasil}
\altaffiltext{9}{Institute of Astronomy and Department of Physics, National Tsing Hua University, 101 Kuang-Fu Rd. Sec. 2, Hsinchu 30013, Taiwan}
\altaffiltext{10}{NSF NOIRLab, 950 N. Cherry Ave., Tucson, AZ 85719, USA}
\altaffiltext{11}{Institute for Computational Cosmology, Department of Physics, Durham University, South Road, Durham DH1 3LE, UK}
\altaffiltext{12}{Department of Astronomy, School of Physics and Astronomy, Shanghai Jiao Tong University, Shanghai 200240, China}
\altaffiltext{13}{University of Texas San Antonio,San Antonio, TX 78249, USA}
\altaffiltext{14}{University of Michigan, 500 S. State Street, Ann Arbor, MI 48109, USA}
\altaffiltext{15}{Lawrence Berkeley National Laboratory, 1 Cyclotron Road, Berkeley, CA 94720, USA}
\altaffiltext{16}{Department of Physics, Boston University, 590 Commonwealth Avenue, Boston, MA 02215 USA}
\altaffiltext{17}{Dipartimento di Fisica ``Aldo Pontremoli'', Universit\`a degli Studi di Milano, Via Celoria 16, I-20133 Milano, Italy}
\altaffiltext{18}{INAF-Osservatorio Astronomico di Brera, Via Brera 28, 20122 Milano, Italy}
\altaffiltext{19}{Department of Physics \& Astronomy, University College London, Gower Street, London, WC1E 6BT, UK}
\altaffiltext{20}{Instituto de F\'{\i}sica, Universidad Nacional Aut\'{o}noma de M\'{e}xico,  Circuito de la Investigaci\'{o}n Cient\'{\i}fica, Ciudad Universitaria, Cd. de M\'{e}xico  C.~P.~04510,  M\'{e}xico}
\altaffiltext{21}{Department of Astronomy, San Diego State University, 5500 Campanile Drive, San Diego, CA 92182, USA}
\altaffiltext{22}{Department of Physics \& Astronomy and Pittsburgh Particle Physics, Astrophysics, and Cosmology Center (PITT PACC), University of Pittsburgh, 3941 O'Hara Street, Pittsburgh, PA 15260, USA}
\altaffiltext{23}{Space Sciences Laboratory, University of California, Berkeley, 7 Gauss Way, Berkeley, CA  94720, USA}
\altaffiltext{24}{University of California, Berkeley, 110 Sproul Hall \#5800 Berkeley, CA 94720, USA}
\altaffiltext{25}{Institut de F\'{i}sica d’Altes Energies (IFAE), The Barcelona Institute of Science and Technology, Edifici Cn, Campus UAB, 08193, Bellaterra (Barcelona), Spain}
\altaffiltext{26}{Departamento de F\'isica, Universidad de los Andes, Cra. 1 No. 18A-10, Edificio Ip, CP 111711, Bogot\'a, Colombia}
\altaffiltext{27}{Observatorio Astron\'omico, Universidad de los Andes, Cra. 1 No. 18A-10, Edificio H, CP 111711 Bogot\'a, Colombia}
\altaffiltext{28}{Institut d'Estudis Espacials de Catalunya (IEEC), c/ Esteve Terradas 1, Edifici RDIT, Campus PMT-UPC, 08860 Castelldefels, Spain}
\altaffiltext{29}{Institute of Cosmology and Gravitation, University of Portsmouth, Dennis Sciama Building, Portsmouth, PO1 3FX, UK}
\altaffiltext{30}{Institute of Space Sciences, ICE-CSIC, Campus UAB, Carrer de Can Magrans s/n, 08913 Bellaterra, Barcelona, Spain}
\altaffiltext{31}{Fermi National Accelerator Laboratory, PO Box 500, Batavia, IL 60510, USA}
\altaffiltext{32}{Center for Cosmology and AstroParticle Physics, The Ohio State University, 191 West Woodruff Avenue, Columbus, OH 43210, USA}
\altaffiltext{33}{Department of Physics, The Ohio State University, 191 West Woodruff Avenue, Columbus, OH 43210, USA}
\altaffiltext{34}{The Ohio State University, Columbus, 43210 OH, USA}
\altaffiltext{35}{Department of Physics, Southern Methodist University, 3215 Daniel Avenue, Dallas, TX 75275, USA}
\altaffiltext{36}{Department of Physics and Astronomy, University of California, Irvine, 92697, USA}
\altaffiltext{37}{Sorbonne Universit\'{e}, CNRS/IN2P3, Laboratoire de Physique Nucl\'{e}aire et de Hautes Energies (LPNHE), FR-75005 Paris, France}
\altaffiltext{38}{Department of Astronomy and Astrophysics, UCO/Lick Observatory, University of California, 1156 High Street, Santa Cruz, CA 95064, USA}
\altaffiltext{39}{Department of Astronomy and Astrophysics, University of California, Santa Cruz, 1156 High Street, Santa Cruz, CA 95065, USA}
\altaffiltext{40}{Departament de F\'{i}sica, Serra H\'{u}nter, Universitat Aut\`{o}noma de Barcelona, 08193 Bellaterra (Barcelona), Spain}
\altaffiltext{41}{Instituci\'{o} Catalana de Recerca i Estudis Avan\c{c}ats, Passeig de Llu\'{\i}s Companys, 23, 08010 Barcelona, Spain}
\altaffiltext{42}{Department of Physics and Astronomy, Siena College, 515 Loudon Road, Loudonville, NY 12211, USA}
\altaffiltext{43}{IRFU, CEA, Universit\'{e} Paris-Saclay, F-91191 Gif-sur-Yvette, France}
\altaffiltext{44}{Department of Physics and Astronomy, University of Waterloo, 200 University Ave W, Waterloo, ON N2L 3G1, Canada}
\altaffiltext{45}{Perimeter Institute for Theoretical Physics, 31 Caroline St. North, Waterloo, ON N2L 2Y5, Canada}
\altaffiltext{46}{Waterloo Centre for Astrophysics, University of Waterloo, 200 University Ave W, Waterloo, ON N2L 3G1, Canada}
\altaffiltext{47}{Instituto de Astrof\'{i}sica de Andaluc\'{i}a (CSIC), Glorieta de la Astronom\'{i}a, s/n, E-18008 Granada, Spain}
\altaffiltext{48}{Departament de F\'isica, EEBE, Universitat Polit\`ecnica de Catalunya, c/Eduard Maristany 10, 08930 Barcelona, Spain}
\altaffiltext{49}{Department of Physics and Astronomy, Sejong University, 209 Neungdong-ro, Gwangjin-gu, Seoul 05006, Republic of Korea}
\altaffiltext{50}{CIEMAT, Avenida Complutense 40, E-28040 Madrid, Spain}
\altaffiltext{51}{Space Telescope Science Institute, 3700 San Martin Drive, Baltimore, MD 21218, USA}
\altaffiltext{52}{Department of Physics \& Astronomy, Ohio University, 139 University Terrace, Athens, OH 45701, USA}
\altaffiltext{53}{Centre for Advanced Instrumentation, Department of Physics, Durham University, South Road, Durham DH1 3LE, UK}
\altaffiltext{54}{National Astronomical Observatories, Chinese Academy of Sciences, A20 Datun Road, Chaoyang District, Beijing, 100101, P.~R.~China}



\begin{abstract}
In this paper we present the stellar Value-Added Catalogue (VAC) based on the DESI Data Release 1. 
This VAC contains stellar parameter, abundance and radial velocity measurements for more than 4 million stars. It also contains, for the first time, measurements from individual epochs for more than a million stars with at least two observations.
The main contribution to the catalogue comes from the bright program of the main survey, which includes $\sim $2.5 million stars, and the backup program, which includes $\sim $ 1 million stars. 
The combined magnitude range for the stars in the catalogue extends from {\it Gaia} G $\sim 12$ to G $\sim 21$. For the magnitude range $17.5<G<21$ this catalogue represents a factor of 10 increase in the number of stars with radial velocity and abundance measurements compared to existing surveys.
Despite DESI's resolution (R $\sim 2500-5000$), the median radial velocity {uncertainty} for stars in the catalogue is better than 1\,km\,s$^{-1}$.
The stellar parameters and abundances of stars in DESI are measured by two independent pipelines, and after applying a temperature-dependent calibration, [Fe/H] abundances of high signal-to-noise stars are accurate to better than $\sim$ 0.1\,dex when compared to high-resolution surveys.  
The catalogue probes different Galactic components including a particularly large number of distant stars:  tens of thousands of stars further than 10\,kpc, and thousands further than 50\,kpc. The catalogue also contains several thousand extremely metal-poor stars with ${\rm [Fe/H]}<-3$.
The released sample of stars includes measurements for thousands of stars that are members of dwarf galaxies, open and  globular clusters as well as members of several dozen stellar streams.
The next public DESI data release is expected in {less than} two years and will contain three times as many stars as DR1.
\end{abstract}

\section{Introduction}

The Dark Energy Spectroscopic Instrument (DESI) is a state-of-the-art, highly multiplexed spectroscopic facility mounted on the Mayall 4-meter telescope at the Kitt Peak National Observatory \citep{FiberSystem.Poppett.2024,Corrector.Miller.2023,FocalPlane.Silber.2023}. DESI's primary mission is to map the three-dimensional structure of the Universe to investigate dark energy through measurements of baryon acoustic oscillations (BAO) and redshift-space distortions \citep{DESI2016a.Science,DESI2016b.Instr}. Following commissioning and Survey Validation (SV) phases from late 2019 to early 2021, DESI started its five-year survey in May 2021, now in its fourth year of operations.

In parallel to cosmology, DESI conducts the Milky Way Survey (MWS) program in bright observing conditions, unsuitable for faint extragalactic targets, to obtain extensive spectroscopy of stellar populations in the Galaxy. The MWS was designed to observe at least seven million stars down to a limiting magnitude of $r \approx 19$, providing radial velocities, stellar atmospheric parameters, and chemical abundances \citep{Cooper2023}. Complementing this primary bright-time program, the backup program (Milky Way Backup Program, or MWBP) exploits even less favourable observing conditions, such as twilight and poor weather, significantly extending both the magnitude range and sky coverage of the survey, producing spectra for several million additional stars, greatly augmenting the primary MWS data \citep{backup_paper}.

The first data set from DESI was released publicly as the Early Data Release (EDR) \citep{DESI_EDR_2024}. Those data were based on survey validation \citep{DESI_SV_PAPER} observations and amounted to some 1.5 million spectra,  including about 400,000 stellar spectra. The analysis of the stellar subset was presented in \citet{Koposov2024_EDR_VAC}, \citetalias{Koposov2024_EDR_VAC} hereafter, showing that DESI is capable of measuring radial velocities accurate to $\sim 1$\,km\,s$^{-1}$ and chemical abundances to $\sim 0.15$\,dex. These data have already been used in several studies \citep{Manser2024,Valluri2025}.

The DESI collaboration recently made publicly available the Data Release 1 (DR1) \citep{DESI_DR1_2025} containing more than 18 million redshifts and approximately four million stellar spectra observed during the first 13 months of main survey operations. These data have been used for multiple cosmological analyses, such  as BAO and clustering measurements using different tracers leading to new constraints on cosmological parameters \citep{DESI2024.VII.KP7B}.
This paper presents the analysis of the stellar subset of the DESI DR1 and an associated Value-Added Catalogue (VAC). In the paper, we highlight the improvements over previous releases and provide guidance on the use of this catalogue. 
The paper is structured as follows. In Section~\ref{sec:data} we describe the key processing steps and improvements that went into the Value-Added Catalogue. In Section~\ref{sec:content} we summarise the contents of the Catalogue. In Section~\ref{sec:validation} we validate  the abundance and velocity measurements. In Section~\ref{sec:discussion} we discuss the possible scientific uses of this dataset as well as compare it to other existing large spectroscopic surveys.

\setcounter{footnote}{0} 
\section{Data}
\label{sec:data}
In this section we describe the DESI data released with DR1 and additional steps undertaken by the Milky Way Survey group to analyse stellar spectra.
We start with a brief reminder of the DESI instrument and structure of the DESI data. 

The DESI instrument is a fiber-fed spectrograph that is capable of observing 5,000 spectra simultaneously over an 8-square-degree field-of-view. The DESI wavelength coverage extends from 3600\,\AA\ in the blue to 9800\,\AA\ in the red. The spectra of each object are split by the dichroic beam splitter into three spectrographs that observe the wavelength ranges from 3600\,\AA\ to 5930\,\AA\ (B arm), 5600\,\AA\ to 8000\,\AA\ (R arm), and 7470\,\AA\ to 9800\,\AA\ (Z arm). The average resolution of the spectra varies with wavelength from $R\sim 2000$ at 3600\,\AA\ to R$\sim 5000$ at 9800\,\AA. All the DESI spectra are extracted at fixed sampling of 0.8\,\AA\ per pixel in the barycentric frame using spectral perfectionism framework \citep{bolton10}, and are flux-calibrated using spectra of stars observed in each exposure.

DESI data are organised by surveys and programs. Table \ref{tab:numbers} gives the list of surveys and programs included in the DR1 together with the number of spectra analysed for this Value-Added Catalogue. The surveys {\tt sv1}, {\tt sv2}, {\tt sv3} correspond to the survey validation programs \citep{DESI_SV_PAPER}, and survey {\tt main} is the main DESI survey.  The survey {\tt special} corresponds to various dedicated pointings and observations, such as observations of M31 used in \citet{Dey2023}. 
The surveys themselves consist of several programs, such as {\tt dark}, {\tt bright}, and {\tt backup}. The programs are completely independent of each other, have different target selection algorithms, and are triggered under different observing conditions. The dark program is executed in the best conditions. The bright program is executed in less-optimal  conditions, such as when the moon is up, while the backup program is conducted under conditions that are below those acceptable for the bright program.  The formal separation of conditions into backup, bright and dark is done using the so-called survey speed, which is determined from the seeing, sky background, transparency, and airmass at the time of observations \citep{Schlafly2023}. 
The dark program includes the main cosmological DESI survey. The bright program includes the Bright Galaxy Survey (BGS) \citep{BGS.TS.Hahn.2023} and the Milky Way Survey (MWS) \citep{Cooper2023}.
The backup program is a separate survey itself, described in detail in \citet{backup_paper}, and is devoted to observations of bright stars $12\lesssim G \lesssim 19$ with an extended footprint compared to the MWS survey. 

\subsection{DR1}

\begin{table}
    \centering
    \begin{tabular}{ccccc}
    Survey & Program & Number of  & Number of \\
     &  &   objects  & stars\\
    \hline
    cmx & other &  736 & 624\\
special & backup &  35,068 & 32,358  \\
special & bright &  18,568 & 12,723  \\
special & dark &  2,275 & 794  \\
special & other &  46,461 & 10,356  \\
sv1 & backup &  61,043 & 57,535  \\
sv1 & bright &  45,768 & 42,318 \\
sv1 & dark &  41,980 & 27,166  \\
sv1 & other &  82,875 & 60,156  \\
sv2 & backup &  2,953 & 2,845  \\
sv2 & bright &  7,239 & 7,009  \\
sv2 & dark &  3,942 & 3,430  \\
sv3 & backup &  83,886 & 82,429  \\
sv3 & bright &  236,550 & 215,646 \\
sv3 & dark &  57,282 & 18,176 \\
\hline
main & backup &  1,218,152 & 1,196,686  \\
main & bright &  3,070,120 & 2,555,246  \\
main & dark &  1,357,709 & 420,750  \\
    \hline
    \end{tabular}
    \caption{The surveys and programs that are part of the DESI DR1 together with the total number of objects with measurements from coadded DESI spectra included in this VAC. The number of objects in the VAC classified as stars by {\tt Redrock} is given in the last column.}
    \label{tab:numbers}
\end{table}

In this paper we focus on the analysis of the DESI Data Release 1 (DR1). DR1 is described in detail in \citet{DESI_DR1_2025}, and we only briefly summarise its contents. It is based on 13 months of DESI observations, including both the reprocessing of the early survey validation data and data from the main DESI survey. The data processing done by the main DESI pipeline and target selection of the survey are described in \citet{guy23,Myers2023}. \citet{DESI_DR1_2025} describes both the DR1 specifics of the DESI pipeline as well as the main official DESI data products released with the DR1. Here we briefly summarise some aspects relevant for this VAC. One of the main products of the DESI pipeline are spectra that are coadded within any given survey and program. The spectra are then grouped by survey, program and HEALPIX \citep{gorski05a} pixels on the sky. 

DESI spectra are flux-calibrated and are in the Solar System's barycentric velocity frame.  All the spectra in DESI have been  analysed by the {\tt Redrock} software \citep{Redrock,Redrock_Anand2024}, which provides spectroscopic classifications and redshifts. In addition to the coadded spectra, DESI provides single-epoch spectra for every target, grouped in a similar fashion to the coadded spectra, i.e., by survey, program and HEALPIX.

The coaddition of spectra within a given survey and program occurs on the basis of the DESI target identifier {\tt TARGETID} described in \citet{Myers2023}, which guarantees that different objects have different identifiers (but the same object may have different {\tt TARGETID}). As mentioned before, the coaddition of spectra happens only within the survey and program, i.e. for  objects that have been observed in multiple surveys and programs, their observations are not coadded together.

\subsection{MWS processing}

The spectra processed by the DESI pipeline and {\tt Redrock} are subsequently analysed by two MWS pipelines: the RVS pipeline and the SP pipeline. They are based on two codes, {\tt RVSpecFit} \citep{rvspecfit} and {\tt FERRE} \citep{2006ApJ...636..804A}. The application of these codes to DESI data is described in \citet{Cooper2023} and \citetalias{Koposov2024_EDR_VAC}, while here we mostly focus on aspects that differ from what was described before. 

The MWS pipelines do not analyse every spectrum observed by DESI and processed by the DESI pipeline. They instead focus on spectra that are of high enough quality and that could belong to stars irrespective of the survey and program they have been observed in. 
Specifically for this data release, only spectra satisfying all of the following requirements have been processed:\footnote{Here we rely on various DESI processing columns defined in \citet{guy23} and \url{https://desidatamodel.readthedocs.io/en/latest/} } 

\begin{itemize}
    \item Good quality, as indicated by {\tt FIBERSTATUS}=0 or {\tt COADD\_FIBERSTATUS}=0.
    \item A signal-to-noise in at least one of the three arms of the instrument larger than 2.
    \item The object type associated with the fiber is not sky, namely ({\tt OBJTYPE}$\neq${\tt SKY}).
\end{itemize}

In addition, to be processed by the MWS pipelines, a DESI spectrum must satisfy at least one of the following requirements (see \citealt{Myers2023} for a detailed definition of the targeting masks):

\begin{itemize}
    \item The object is an MWS target (i.e. has {\tt MWS\_ANY} targeting flag).
    \item The object is a flux standard star (i.e. it has any {\tt STD\_} targeting mask set).
    \item The object is a secondary target (i.e. it has the {\tt SCND\_ANY} flag set). 
    \item The object is classified as a star by {\tt Redrock}.
    \item The object has an absolute value of the radial velocity measured by {\tt Redrock} below 1500\,km\,s$^{-1}$.
\end{itemize}

This means that for example MWS targets that ended up being quasars or galaxies, or targets of any kind spectroscopically classified as galaxies at redshifts $z\lesssim 0.005$ are all analysed by the MWS pipelines and are included in the VAC. Similarly, objects that were targeted as galaxies or quasars for the main DESI cosmological survey but ended up being stars are also part of the VAC.

The logic described above is applied to select coadded DESI spectra for processing by RVS and SP pipelines. However, in this data release, we also incorporate the analysis of single-epoch observations by the RVS pipeline. 
To select single-epoch spectra that are fitted, the same set of conditions as described above is adopted. The only difference is that we still use the {\tt Redrock} classification and redshift from the coadded spectra. 

\subsection{Stellar analysis pipelines}
The RVS and SP pipelines are described in detail in \citetalias{Koposov2024_EDR_VAC}. However, several improvements were implemented for the DR1 stellar VAC, primarily to the RVS pipeline. We discuss these changes below.

\subsubsection{RVS pipeline}

The RVS pipeline is powered by the {\tt RVSpecFit} code which forward-models  DESI spectra by this equation:

\begin{equation}
\begin{array}{c}
M(\lambda| v, \phi) = (\sum_i a_i P_i(\lambda)) \times S(\lambda \cdot ( 1 + v/c)| \phi) 
\end{array}
\label{eq:rvspecfit}
\end{equation}

\noindent (see \citealt{koposov2011}), where $P_i(\lambda)$ are polynomials or other basis functions of wavelength, $v$ is the radial velocity, $a_i$ fitted constants and $S(\lambda|\phi)$ is the interpolated stellar spectrum model for the stellar parameters $\phi$. This and the previous value-added catalogue for DESI EDR were based on the PHOENIX stellar library  \citep{husser2013} with the $\phi$ parameters being surface gravity, effective temperature,  metallicity and alpha-element abundance: $\log g$, $T_{\rm eff}$, [Fe/H], ${\rm [\alpha/Fe]}$. For the EDR catalogue presented in \citetalias{Koposov2024_EDR_VAC} the code used multi-linear interpolation in the space of [Fe/H], ${\rm [\alpha/Fe]}$, $\log g$, $T_{\rm eff}$ to construct interpolated spectra $S(\lambda|\phi)$. The discontinuity of derivatives at the grid nodes led to concentration of measured  stellar parameters there (see Figure 9 of \citetalias{Koposov2024_EDR_VAC}).

For this release we use an updated version of the {\tt RVSpecFit} code, where the stellar spectrum model $S(\lambda| \phi)$ is constructed using a neural network emulator. This was done to address multiple problems highlighted in \citetalias{Koposov2024_EDR_VAC}. We use a feed-forward fully-connected neural network with four input parameters $\log g$, $T_{\rm eff}$, ${\rm [Fe/H]}$,  ${\rm{[\alpha/Fe]}}$. There are three 128-d  hidden layers of the network, and the SiLU (Sigmoid Linear Unit) is used as a non-linearity. These layers of the network return a 200-d vector which is then passed through a final fully linear layer transforming 200-d vector into a DESI-like spectrum (i.e., spectrum with several thousands pixels, $\sim$ 0.4\,\AA\ sampling and spanning the wavelength range of DESI data). The neural network was implemented using the {\tt pytorch} framework and trained  on PHOENIX spectra using the {\tt  Adam} optimiser \citep{kingma2017adammethodstochasticoptimization} with the {\tt ReduceLROnPlateau} scheduler.

Since {\tt RVSpecFit} models multiple arms of the DESI spectra simultaneously, three different neural network emulators were constructed, one for each of the arms of the spectrograph.
The neural network emulators were trained using PHOENIX synthetic spectra convolved with a Gaussian kernel to DESI resolution (assuming a fixed FWHM of 1.55\,\AA\ for the line spread function in the B and R arms, and a FWHM of 1.8\,\AA\ for the Z arm). While the trained neural network emulator cannot reproduce PHOENIX spectra exactly, we have found that the median absolute deviation of the emulated spectra with respect to the actual PHOENIX models is less than 0.5\% (0.4\%  for the blue arm, 0.1\% for the R and Z arms). 

Another modification to {RVS} pipeline is an additional warning flag that is set in the {\tt RVS\_WARN} bitmask column. The fourth bit (corresponding to value of 8) of this bitmask is set if some of the stellar parameters are close to or beyond  the stellar grid edges. Specifically, if the [Fe/H] is within 0.01 dex of $-4$ or $1$, and if the $T_{eff}$ is within 10\,K of 2,300\,K or 15,000\,K the fourth bit of {\tt RVS\_WARN} is enabled. We recommend that when analysing stellar parameters from {RVS}, only stars with {\tt RVS\_WARN}=0 should be used, but at the same time many stars that have well measured radial velocities and are well fit by RVS have this warning flag.
    
The change of the interpolator is the main improvement to the RVS pipeline with respect to the version used for the construction of the EDR VAC.
Other small improvements and bug fixes were also adopted, such as addressing the issues with stars with high rotational velocity, $v \sin i$, described in the EDR paper, and improving the uncertainties of stellar parameters.

\subsubsection{SP pipeline}

The SP pipeline is powered by {\tt FERRE} \citep{Allende-prieto23a} code that is designed to measure stellar atmospheric parameters  and individual elemental abundances by fitting continuum normalised spectra.
There were no substantial changes in the SP pipeline, other than 
an increase in the number of elements for which abundances are derived, from four to ten.
It now includes: Fe, Ca, C, Mg, Si, Na, Cr, Ni, Al, and Ti, reported in that order in the {\tt ELEM} array within the SPTAB extension of the summary files.
As with the EDR data, the results that were fitted with Kurucz model atmospheres \citep{kurucz2005} 
should be used, avoiding those fitted with PHOENIX models. This means we suggest to use {\tt BESTGRID} $\neq$ `s\_rdesi1' when working with SP measurements. Despite the efforts to include more chemical elements, we deem only the elemental abundances of Mg, Ca, and Fe to be reliable.

\section{Contents of the release}
\label{sec:content}

In this section we provide a brief overview of the contents of the catalogue, highlighting the aspects that are different from 
the EDR VAC.

The file structure of the catalogue and column names have  been described in \citetalias{Koposov2024_EDR_VAC}, and are mostly identical in this VAC. The data model is available online.\footnote{\url{https://desi-mws-dr1-datamodel.readthedocs.io/en/latest/}}

Similar to the EDR VAC, this release includes measurements by RVS and SP pipelines of coadded spectra from different surveys and programs. The release also contains a merged catalogue (named {\tt mwsall-pix-iron.fits}) that contains the combination of the RVS and SP measurements across all the surveys, targeting information, and the crossmatch with {\it Gaia} DR3 source catalogue \citep{gaia_main,gaia_edr3,gaia_dr3}. This combined catalogue is  probably the most useful to work with for researchers interested in Milky Way science with DESI. We provide a short reminder of the structure of this file in the Appendix~\ref{sec:appendix_mwsall}, and also release a tutorial on how to use the catalogue.\footnote{\url{https://github.com/desimilkyway/dr1_tutorials/}}
The validation of the data quality of the catalogue is discussed in Section~\ref{sec:validation}, but the full list of known issues affecting the catalogue that users need to be aware of is provided in the Appendix~\ref{sec:known_issues}.

A new component of this release of the Value-Added Catalogue is the RVS pipeline measurements from single-epoch spectra. Those are provided both for individual HEALPIX pixels within each survey and program, and in a combined form of all single-epoch measurements within a survey and program. Similar to catalogues of measurements from coadded spectra, the combined tables include the crossmatch with {\it Gaia} and targeting information.

In the sections below we provide some key statistics for the catalogue constructed from co-added spectra and single-epoch spectra.

\subsection{Co-added spectra analysis}

Table~\ref{tab:numbers} gives numbers of objects in the VAC with measurements from coadded DESI spectra, grouped by  different surveys and programs. The table also provides the number of objects that were classified by {\tt Redrock} as stars. 
The table shows that the majority of sources in the release ($\sim2.5$ million with a star classification by {\tt Redrock}) originate from the bright program of the main survey. The table also shows that a large number of sources come from the backup program ($\sim 1$ million)  and dark program ($\sim 0.5$ million). 
For the dark program in particular, many objects in the catalogue are not classified as stars by {\tt Redrock} ($\sim $ 70\%), while across other programs the fraction of non-stars  is below 20\%. The reason for this is mostly because the dark program contains many secondary targets that are not stars.
The total number of coadded stellar spectra analysed in this VAC across all programs and surveys is 6,372,607. 4,746,246 of those are classified as a star by {\tt Redrock}.

The numbers in Table~\ref{tab:numbers} may include duplicates, as objects observed in multiple programs are counted for each. The combined catalogue of measurements (the {\tt mwsall} file) for the stellar VAC includes measurements from all the surveys and programs, and in the case of objects observed in more than one program and survey, the observation with the highest signal-to-noise (in the R arm) is marked as primary (this can be identified using the {\tt PRIMARY} column in the table).\footnote{The identification of primary observation happens not only for sources that have the same TARGETID in multiple programs and surveys, but also for sources with different TARGETID that are found to be within 0.5 arc-seconds of each other at epoch of J2021.0. To calculate the source position at J2021 we use the proper motions from the targeting catalogue {\tt FIBERMAP}.}  The number of primary (unique) objects in the combined catalogue is 5,930,124. 4,378,110 of those are classified as a star by {\tt Redrock}

\subsubsection{Survey footprint}

\begin{figure*}
    \centering
    \includegraphics{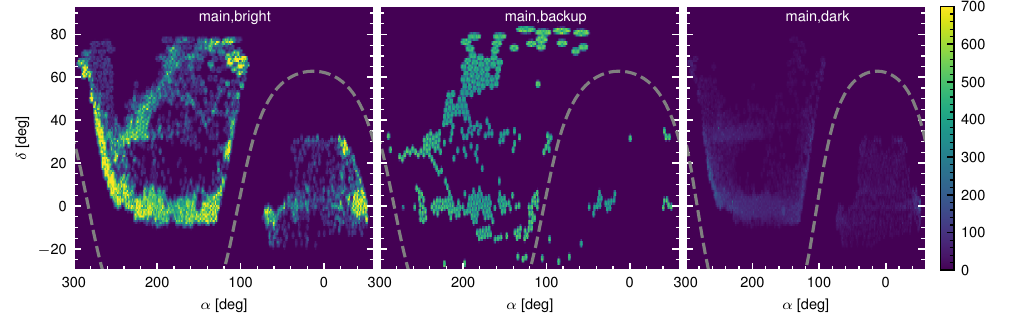}
    \caption{Density of objects on the sky that are included in the VAC, were observed in the main survey and are classified as stars. 
    Each panel shows a different program of the main survey. 
    The units for the density are in objects per square degree. The dashed line shows the Galactic plane.}
    \label{fig:density}
\end{figure*}

The footprint of the data included in this VAC is dramatically larger (by a factor of 7) than the one released with the EDR. Figure~\ref{fig:density} shows the density of objects on the sky classified as stars in the VAC for three programs of the main survey (see Figure 1 of \citetalias{Koposov2024_EDR_VAC} for comparison with EDR). 

The backup program footprint extends further than the other programs and beyond the DR9 footprint of the DECaLS survey \citep{LS.Overview.Dey.2019}, as the backup targets are selected using {\it Gaia} rather than DECaLS photometry \citep[see ][for more details]{backup_paper}.  The figure shows that the density of targets in the catalogue varies dramatically across the sky for dark and bright programs, because of the different degree of progress of the survey in different areas of the sky. The reason for this is that the bright program requires five passes across the survey footprint, while for the dark program seven passes are required \citep[see ][]{Schlafly2023}. The backup program is done using a single pass over the footprint, which results in a more uniform density of objects in the catalogue seen in Figure~\ref{fig:density} (roughly equal to DESI's fiber density of  5000 objects per 8 square degree of field of view). 
The dark program has a significantly lower density of stellar targets compared to the bright or backup programs, as the majority of observed stars in the dark program are only observed as flux calibration standards (which typically occupy some 100 fibres out of 5000 -- see \citealt{guy23}).

\begin{figure*}
    \centering
    \includegraphics{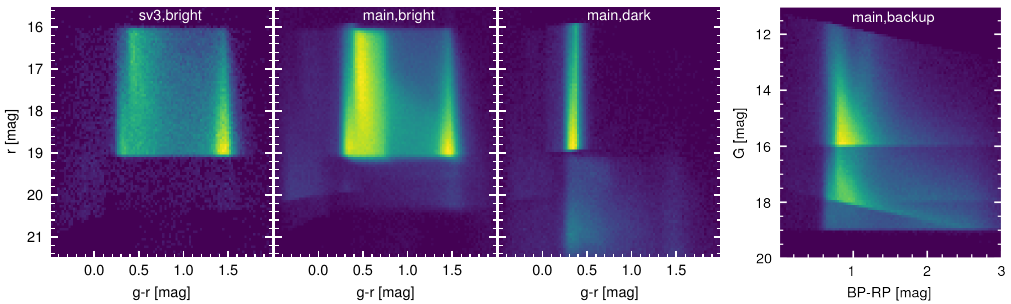}
    \caption{{\it Left 3 panels:} The colour-magnitude distribution in DECaLS photometric bands for objects included in the VAC. Different panels (from left to right) show objects from the bright program of the {\tt sv3} survey, dark and bright programs of the main survey respectively.  {\it Right panel:} The colour-magnitude distribution of targets in the backup program included in the VAC in {\it Gaia} photometric bands. All the panels only show objects classified as a star by {\tt Redrock}. No extinction correction was applied to the magnitudes.}
    \label{fig:cmd}
\end{figure*}

\begin{figure}
    \centering
    \includegraphics{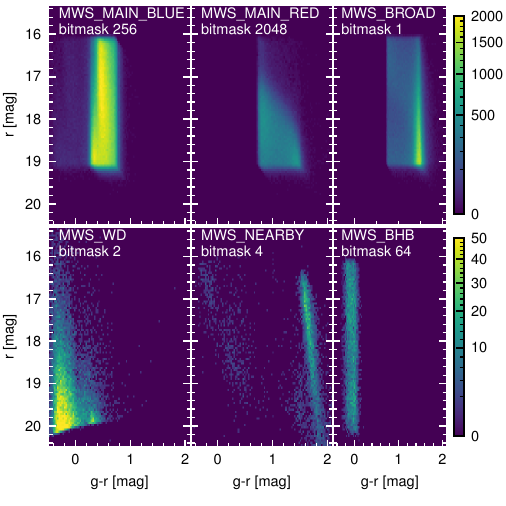}
    \caption{The colour-magnitude distribution for different target classes in the VAC observed in the bright program of the main survey. The description of the selections is provided in \citet{Cooper2023}. The most numerous targets are the {\tt MWS\_MAIN\_BLUE} and {\tt MWS\_MAIN\_RED}, where the former selects  blue sources in the $16\lesssim r\lesssim19$ magnitude range, the latter targets red sources but also applies parallax and proper motion cuts to prioritise distant giants. The other large target class is {\tt MWS\_BROAD}, which is a magnitude limited selection of any object not selected by {\tt MWS\_MAIN\_BLUE} and {\tt MWS\_MAIN\_RED} categories. The target  boundaries in the plot appear somewhat blurred as selections are applied to dereddened magnitudes, while we show raw magnitudes. The colour-magnitude distributions for high priority targets, such as white dwarfs, stars within 100\,pc and blue horizontal branch stars are shown in the bottom three panels. On each panel we also give the bitmask value that can be used to select these targets using {\tt MWS\_TARGET} column in the catalogue.}
    \label{fig:targ_classes}
\end{figure}

\subsubsection{Survey targets}

As described earlier, the DESI stellar catalogue contains objects selected in various ways in different programs. However the dominant source of stars in it is the Milky Way Survey. The MWS targeting strategy has been described in \citet{Myers2023} and \citet{Cooper2023}, but here we highlight the actual  targets observed and included in the VAC, while focusing on targets observed in the main survey. 
Figure~\ref{fig:cmd} shows the colour-magnitude diagrams for objects classified as stars in three programs of the main survey and the bright program of the survey validation 3 ({\tt sv3}) for comparison.  The magnitudes were not corrected for extinction in this plot.
The colour-magnitude distribution of stars in the bright program of the main survey is mostly showing the target selections of the MWS and is similar to the colour-magnitude distribution in the {\tt sv3} survey, as little was changed in the selection. The distribution shows that the stellar targets in the bright survey are mostly limited to $16\lesssim r \lesssim 19$, with a small fraction of fainter sources. The colour-magnitude diagram for the dark program of the main survey shows that most objects have magnitudes $16\lesssim r\lesssim 19$ similar to the bright program, but they are within a narrow colour range around $g-r\sim$0.3. This reflects the selection of standards for spectro-photometric calibration of DESI. The objects at fainter magnitudes ($r>19$) in the dark program are either high priority MWS targets observed in both the dark and bright surveys, or additional secondary targets \citep[such as faint blue horizontal branch (BHBs), see][]{bystrom24a}. 

The rightmost panel of Figure~\ref{fig:cmd} shows the colour-magnitude distribution of stars in the backup program. This colour-magnitude diagram is shown in {\it Gaia} photometric bands as objects in this program were selected based on {\it Gaia} photometry rather than Legacy Survey photometry.
Sharp features in the colour-magnitude distribution are visible in this panel,
caused by the selection limits for various target classes \citep[see ][for more details]{backup_paper}.  
The targets in the backup survey data extend all the way to {\it Gaia} $G=19$.

Since the majority of objects included in the VAC come from the MWS, in Figure~\ref{fig:targ_classes} we show the colour-magnitude distribution of different MWS target classes observed in the bright program of the main survey. 
The formal definitions of these selections are given in \citet{Cooper2023} and \citet{Myers2023}. The top panels show the distribution of the most numerous  targets such as blue purely photometrically selected targets ({\tt MWS\_MAIN\_BLUE}), red targets  prioritising distant giants ({\tt MWS\_MAIN\_RED}), and a broad magnitude-limited selection of targets ({\tt MWS\_BROAD}) that  were not in the  {\tt MWS\_MAIN\_BLUE} or {\tt MWS\_MAIN\_RED} selections. The {\tt MWS\_MAIN\_BLUE} sample  is dominated by stars around the main sequence turn-off and has a quite uniform magnitude distribution. The {\tt MWS\_MAIN\_RED} sample shows a distribution with a small number of bright objects with $r<18$ and increasing number of fainter objects. The reason for this is that the parallax and proper motion cuts favouring distant giants become less effective at faint magnitudes due to poorer {\it Gaia} astrometric accuracy. The {\tt MWS\_MAIN\_BROAD} sample mostly contains stars with the same colours and magnitudes as the {\tt MWS\_MAIN\_RED} sample. In the bottom three panels of the Figure~\ref{fig:targ_classes} we show higher priority targets such as white dwarfs, BHB stars, and a sample of stars within 100\,pc. 
There are additional categories of MWS targets, such as RR Lyrae and fainter BHB targets that have been observed as secondary targets (and thus can be identified using the {\tt SCND\_TARGET} bitmask rather than {\tt MWS\_TARGET}), that
are not shown in the figure.  

Table~\ref{tab:numbers_classes} provides a summary of the number of MWS  targets observed in different programs of the main survey as well as the number of different MWBP targets. {\tt MWS\_MAIN\_BLUE} is the most dominant target class, and some objects from it have been observed in dark time, because some of the {\tt MWS\_MAIN\_BLUE} targets are also spectrophotometric standards (see Figure~\ref{fig:cmd}). The Table reveals that some of the high priority targets (white dwarfs, blue horizontal branch stars) were observed in dark and bright time. 

While the discussion of the survey selection function is beyond the scope of this paper, we stress that the input target catalogues, as well as their selection criteria for the main survey were published \citep{Myers2023} and are available at \url{https://data.desi.lbl.gov/public/dr1/target/}. Within a given survey and program (e.g. bright program of the main survey), we expect that for objects that are only members of one target class (i.e. {\tt MWS\_MAIN\_BLUE}) the probability of having a DESI spectrum is only a function of the number of DESI passes over given area and possibly local target density, but not any of the source properties such as magnitude, colour or proper motion.

\begin{table}
    \centering
    \begin{tabular}{cc}
    Type & N objects  \\
    \hline
Survey, program: main, dark  & \\ \hline
MWS\_BROAD &  356 \\
MWS\_WD &  8,743 \\
MWS\_NEARBY &  702 \\
MWS\_BHB &  1,869 \\
MWS\_MAIN\_BLUE &  256,206 \\
MWS\_MAIN\_RED &  1,398 \\
\hline 
Survey, program: main, bright  & \\ 
\hline
MWS\_BROAD &  531,420 \\
MWS\_WD &  39,524 \\
MWS\_NEARBY &  8,775 \\
MWS\_BHB &  8,655 \\
MWS\_MAIN\_BLUE &  1,499,825 \\
MWS\_MAIN\_RED &  303,526 \\
\hline 
Survey, program: main, backup  & \\ 
\hline
GAIA\_STD\_FAINT &  162,532 \\
GAIA\_STD\_WD &  1,102 \\
GAIA\_STD\_BRIGHT &  135,833 \\
BACKUP\_GIANT\_LOP &  148,239 \\
BACKUP\_GIANT &  117,882 \\
BACKUP\_BRIGHT &  636,850 \\
BACKUP\_FAINT &  196,848 \\
BACKUP\_VERY\_FAINT &  116,580 \\
    \hline
    \end{tabular}
    \caption{Number of objects in different target classes observed in three programs of the main survey included in this VAC. For bright and dark programs we show MWS target classes~\citep{Cooper2023}, and for the backup program we show backup program specific target classes described in  \citet{backup_paper}. }
    \label{tab:numbers_classes}
\end{table}

\subsection{Single-epoch measurements}

In this section we summarise the products of the analysis of single-epoch observations of DESI spectra with the RVS pipeline (the SP pipeline was not run on single-epoch spectra).   
This is a new data product in DR1 VAC, not previously released for the EDR. 

While repeated observations of targets were not the main goal of the MWS, many objects are in fact observed multiple times across the duration of the survey. There are several reasons for this:
\begin{itemize}
    \item Some stars are flux calibration standards, and any DESI tile covering the same area will tend to repeatedly select these targets.
    \item In the case of multiple survey passes covering the same area, some targets may be re-observed due to the lack of alternative targets with high enough priority.
    \item Some target classes specifically prioritise multiple observations \citep[e.g., {\tt MWS\_BHB}; see ][]{Myers2023}.
   \item When the observation of a given DESI tile does not reach the required effective exposure time (e.g. 180 seconds for the bright program of the main survey), an additional observation will be attempted \citep[see ][for more details]{guy23,Schlafly2023}. This can happen for example in case of observing conditions that become worse throughout the duration of the exposure. That happens particularly frequently with the backup program \citep{backup_paper}.
\end{itemize}

Table~\ref{tab:nrepeats} gives a total number of {\it Gaia} sources in this VAC with more than one DESI observation within different surveys and programs (note that some sources have been observed many times across multiple surveys and programs as well). The table shows that the largest number of repeated observations comes from the backup program.
This is caused by the conditions of observations being poor and DESI tiles in that program often requiring more than one exposure. Around half a million objects have also been observed multiple times in the bright program of the main survey, followed by quarter of a million objects with re-observations in the dark program. The {\tt sv3} survey also provides a significant number of objects with repeated measurements, because of the so-called 1\% survey \citep[see][]{DESI_SV_PAPER}, where a small area of the sky was observed repeatedly to achieve high degree of completeness across all target classes.

The total number of single-epoch spectra measurements for this VAC is 10,012,925. This includes around 4 million spectra for objects observed only once, where the spectrum is mostly identical to the coadded spectrum.\footnote{The coadded single epoch spectra are not exactly identical to single epoch spectra due to an identified resolution matrix and masking bug that affects a tiny fraction of pixels in DESI spectra, see Appendix E in \citet{DESI_DR1_2025}}

In Figure~\ref{fig:nrepeats} we show the distribution of the number of repeated DESI observations for different {\it Gaia} objects, grouped by the maximum interval between earliest and latest observations. The figure shows that most objects that were observed more than once have a small number (2 to 3) of repeated observations, but there is an extended tail to many tens of observations, with more than $70,000$  
objects having ten or more repeats. Most of the objects with a large number of repeated observations have been observed in the survey validation program. The figure also shows that while most objects with repeated observations were observed within a short time span, a substantial fraction of objects have observations separated by a few months. Note that this VAC is based on the dataset obtained only over 13 months of the main survey and 5 months of survey validation. Since then DESI has been observing for three years, so the future data release will have  a significantly larger baseline of repeated observations.

\begin{figure}
    \centering
    \includegraphics{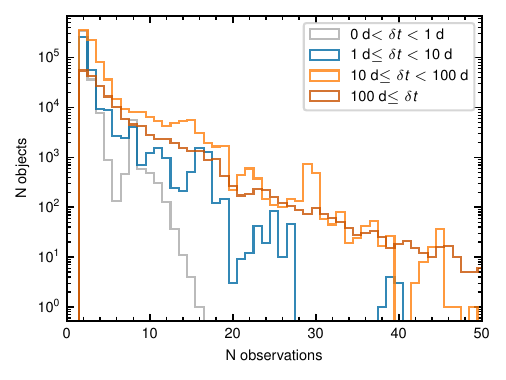}
    \caption{Histogram of the number of repeated observations in DESI DR1 for {\it Gaia} sources in this VAC. 
    Curves of different colours ({grey, blue, green, orange, brown}) show the histograms for subsets of objects where the time separation between the last and first observation in DR1 is less than 1 day, between  1 and 10 days, 10 and 100 and more than 100 days respectively.}
    \label{fig:nrepeats}
\end{figure}

\begin{table}
    \centering
    \begin{tabular}{ccc}
    Survey & Program & Number of objects with repeats \\
    \hline
cmx & other &  527 \\
special & backup &  19,509 \\
special & bright &  11,769 \\
special & dark &  1,052 \\
special & other &  1,798 \\
sv1 & backup &  28,792 \\
sv1 & bright &  41,465 \\
sv1 & dark &  25,815 \\
sv1 & other &  56,897 \\
sv2 & bright &  4,574 \\
sv2 & dark &  1,521 \\
sv3 & backup &  25,950 \\
sv3 & bright &  123,812 \\
sv3 & dark &  17,171 \\
\hline
main & backup &  627,492 \\
main & bright &  443,413 \\
main & dark &  240,100 \\
\hline
    \end{tabular}
    \caption{The number of {\it Gaia} sources with more than one DESI measurement in the Value-Added Catalogue grouped by different surveys and programs.} 
    \label{tab:nrepeats}
\end{table}

\section{Validation}
\label{sec:validation}

In this section we look at the measurements in the catalogue, such as radial velocity, stellar parameters and abundances and compare them with other catalogues. 
We also inspect the radial velocity precision from repeated observations.

We first apply a set of quality cuts to the sample; specifically, we use the recommendations from   \citetalias{Koposov2024_EDR_VAC}:
\begin{itemize}
\item {\tt RR\_SPECTYPE} = STAR
\item {\tt RVS\_WARN} = 0
\end{itemize}
applied to all measurements and {\tt VSINI} < 30\,km\,s$^{-1}$ when using RVS pipeline parameters. 
For measurements from the SP pipeline we apply the selection cut 
{\tt BESTGRID} $\neq$ `s\_rdesi1'.

\subsection{Distribution of stellar parameters and abundances}

\begin{figure}
    \centering
    \includegraphics{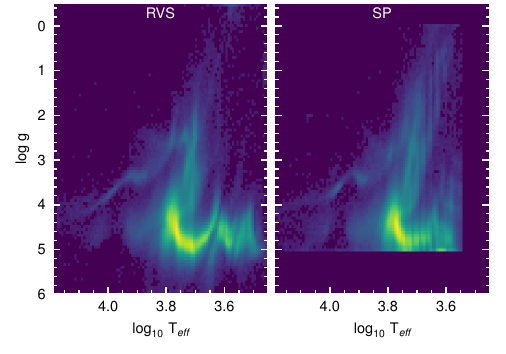}
    \caption{Distribution of the surface gravity and effective temperatures measured by two DESI stellar  pipelines (RVS on the left, SP on the right). The figure only shows stars observed in the main survey with the signal-to-noise in the R arm above 10. }
    \label{fig:loggteff}
\end{figure}

Figure~\ref{fig:loggteff} shows the distribution of surface gravity $\log g$ vs  $T_{\rm eff}$ (Kiel diagram) using measurements from both stellar pipelines. We show the stars observed in the main survey with signal-to-noise in the R arm of the instrument above 10. The main improvement in this distribution compared to the EDR catalogue (c.f.  Figure 9 in \citetalias{Koposov2024_EDR_VAC}) is the disappearance of clustering near the grid-points in $\log g$ and $T_{\rm eff}$ for the RVS catalogue. Otherwise, the distribution shows the expected main sequence, red giant branch as well as horizontal branch. For dwarf stars ($\log g >4$) with lower effective temperatures ($T_{\rm eff}<4500$\,K), the surface gravity is not a smooth function of temperature, similar to what was seen in EDR. The morphology of the diagrams for RVS and SP pipeline is similar, but the SP parameters show some gridding of temperatures for cool stars; also, the RVS red giant branch seems less populated at $\log g <2$. 

\begin{figure}
    \centering
    \includegraphics{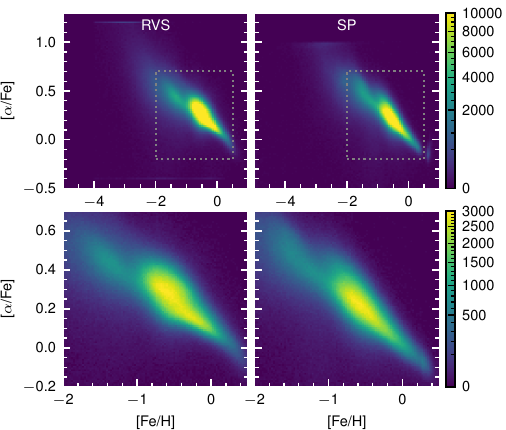}
    \caption{Iron abundance [Fe/H] vs ${\rm [\alpha/Fe]}$ for stars in the main survey as measured by two DESI stellar pipelines. We only include stars with signal-to-noise in the R arm above 10 and with effective temperatures between 4500 and 7000\,K. The bottom panels show a smaller range of [Fe/H] and  ${\rm [\alpha/Fe]}$ corresponding to the grey dashed box in the top panels that is predominantly occupied by the MW disc and Gaia-Sausage-Enceladus stars. The observed high values of $[{\rm \alpha/Fe}]$ for low metallicity stars [Fe/H]$<-2$ are likely unphysical.}
    \label{fig:feh_alphas}
\end{figure}

In Figure~\ref{fig:feh_alphas} we show measured iron abundances [Fe/H] vs ${[\rm \alpha/Fe]}$ from two stellar pipelines. Here we show stars in the main survey with effective temperatures  $4500\ {\rm K}<T_\mathrm{eff}<7000\ {\rm K}$ and signal-to-noise ratio above 10 in the R arm. We again see that the gridding effect from the RVS pipeline has disappeared. The RVS and SP distributions seem to agree well and show some slight bimodality at ${\rm [Fe/H]}\approx -0.5 $,  where thin and thick disk are expected to separate from each other \citep{hayden2017}. The continued rise of $\rm{[\alpha/Fe]}$ at low  metallicities ${\rm [Fe/H]}<-2$ is likely unphysical, since it is not seen in high-resolution measurements of the stellar halo \citep{Naidu2020}.
We also observe a small, likely unphysical, overdensity of stars at the edges of the PHOENIX grid in $\rm{[\alpha/Fe]}$ visible for the RVS measurements at $\rm{[\alpha/Fe]}$ of $-0.4$ and 1.2. For these reasons, we do not recommend using DESI's $\rm{[\alpha/Fe]}$ at low metallicities (below ${\rm [Fe/H]}\lesssim -2$).

\begin{figure}
    \centering
    \includegraphics{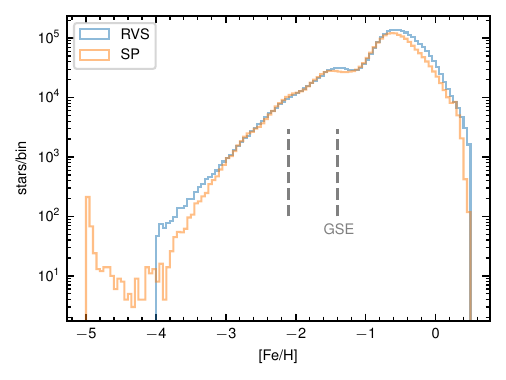}
    \caption{Metallicity distribution of stars observed in the main survey. We show stars with effective temperatures 4500\,K$<T_{\rm eff}<$7000\,K and signal-to-noise above 10 in the R arm. Blue and orange coloured curves show the distribution of [Fe/H] measured by the RVS and SP pipelines respectively. The dashed lines identify several locations where the metallicity distribution flattens, indicative of possible contribution of a dwarf galaxy debris. One such location [Fe/H]$\approx-1.3$ is where the peak of the GSE metallicity distribution is, while the line at [Fe/H]$=-2.1$ likely indicates another accreted system.}
    \label{fig:feh_distr}
\end{figure}

Finally, Figure~\ref{fig:feh_distr} shows the metallicity distribution for stars in the main survey. We use the same selection of stars as for the previous figure. The distributions clearly show the peaks associated with the MW disk, and Gaia Sausage-Enceladus at [Fe/H]$\sim-1.3$ \citep[see, e.g.,][]{Belokurov2018,Helmi2018,Feuillet2020}. We note that although the metallicity distribution shows an almost power-law-like  behaviour from [Fe/H]$=-2$ to lower metallicities, it also shows small "wiggles", for example at [Fe/H]$\approx-2.1$, where the metallicity distribution (as measured by two pipelines) briefly flattens. That could be the result of debris deposition from a relatively massive progenitor with the mean metallicity of $-2.1$ \citep[i.e. something like Thamnos or I'itoi; see ][]{Naidu2020}. Overall the distribution of RVS and SP metallicities show similar behaviour, except near the edges. At high metallicities the RVS sample has more stars than SP (this is mainly caused by the exclusion of SP measurements fitted by PHOENIX models). At low metallicities we can see some differences in 
the SP and RVS distributions of
[Fe/H] mostly due to the fact that the PHOENIX grid used by the RVS pipeline has a limit at [Fe/H]$=-4$, while the SP pipeline has a limit at [Fe/H]$=-5$. We note that the SP metallicities show an excess near the grid edge at [Fe/H]=$-5$ which is not physical. Using the same sample as shown in the figure, we can estimate the total number of extremely metal-poor stars (i.e with [Fe/H]$<-3$) in the bright program of the main survey -- it is $\sim 4000$ if we use RVS abundances, and $\sim $ 3000 with SP abundances. These numbers rise by 2,000 if we include all surveys and programs.

\subsection{Comparison of abundances to external measurements}
\label{sec:feh_compar}
In the previous section we looked at the overall distribution of stellar parameters and abundances. In this section we focus on iron abundance and compare 
our results to those from
other large medium- and high-resolution surveys such as APOGEE \citep{majewski17a},  GALAH  \citep{galah0} and {\it Gaia} RVS \citep{Cropper2018}. A similar comparison to ${\rm [\alpha/Fe]}$ and Ca and Mg abundances is presented in the Appendix~\ref{sec:appendix_alpha_elt_abundances}. We use the data from the DR17 catalogue \citep{apogee_dr17} of APOGEE, the DR4 catalogue \citep{Buder2024} of GALAH, and the DR3 catalogue of stellar parameters from {\it Gaia} RVS \citep{Recioblanco2023}.   
For APOGEE stars we require  {\tt fe\_h\_flag} = 0 and {\tt starflag}=0, for GALAH we require {\tt flag\_SP}=0 and {\tt flag\_fe\_h}=0,  and for {\it Gaia} we require the first 13 flags of {\tt flags\_gspspec} to be zero. We also include in our comparison a compilation of measurements of metal-poor stars from the SAGA database \citep{SAGA}.\footnote{Using version from April 7, 2021. Available at \url{http://sagadatabase.jp/Data/txt_table_all.tsv}.}

In the cases of GALAH, APOGEE and {\it Gaia} we use the {\it Gaia} DR3 {\tt source\_id} from the DESI catalogue to match the sources. Since the SAGA catalogue lacked accurate or complete celestial coordinates, we retrieved these coordinates from the Simbad \citep{Simbad} database using the {\tt astroquery} package based on the {\tt Object} identifiers, and then cross-matched with DESI using a 1$\arcsec$ radius. Stars without valid Simbad matches were excluded from the comparison.

 For the rest of this subsection, in addition to the DESI selections listed in the beginning of Section~\ref{sec:validation}, we also require the signal-to-noise to be above 10 for the instrument's R arm and require that DESI's uncertainty on [Fe/H] is lower than 0.1 dex.   The typical uncertainty of the abundances in the comparison sample from GALAH is $0.05$\,dex, 0.01 for APOGEE sample and 0.28 for {\it Gaia} sample.
 
Figure~\ref{fig:feh_compar1} compares the metallicities measured by the two DESI pipelines with those of other surveys. The left (right) columns refer to the RVS (SP) pipeline. Different rows show different surveys. We notice that there is a good correlation between DESI's measurements and external measurements, however there are some clearly identifiable issues. The comparison with APOGEE clearly shows two sequences, one consistent with a small positive offset, and another with a negative offset, shared by the two pipelines. The SAGA sample shows a systematic bias in metallicities similarly shared by the two pipelines. The issue specific to the RVS pipeline seems to be a larger spread above the one-to-one line indicative of [Fe/H]$_\mathrm{DESI,RVS} >$[Fe/H]$_\mathrm{external}$.

Given the significant structure in Figure~\ref{fig:feh_compar1}, it  is helpful to look at the residuals of DESI's metallicities with respect to other surveys, i.e. [Fe/H]$_\mathrm{DESI}$-[Fe/H]$_\mathrm{external}$. These are shown in Figure~\ref{fig:feh_compar2}, where we also compare to the  abundances from {\it Gaia} DR3 RVS spectra \citep{Recioblanco2023}.  In each panel we show the distribution of residuals to different surveys, and the RVS and SP results are shown by curves of different colour. In each panel we also provide the median and the 16/84-th percentiles of the distributions. The figure shows that raw DESI abundances show deviations of $\sim 0.1-0.25$\,dex when compared to GALAH and APOGEE. The RVS and SP seem to show similar biases with respect to GALAH and APOGEE, but RVS seems to perform noticeably worse than SP on the {\it Gaia} set. The figure also confirms the bias of 0.2 dex of DESI metallicities from both pipelines for the metal-poor SAGA sample. It also shows more clearly that there is a bimodality in the metallicity offsets of DESI, such that some objects show positive  and some show negative metallicity offsets. This bimodality is particularly visible with the APOGEE and {\it Gaia} samples.

\subsubsection{Calibration of [Fe/H] abundances}
\label{sec:feh_calib}
\begin{figure}
    \centering
    \includegraphics{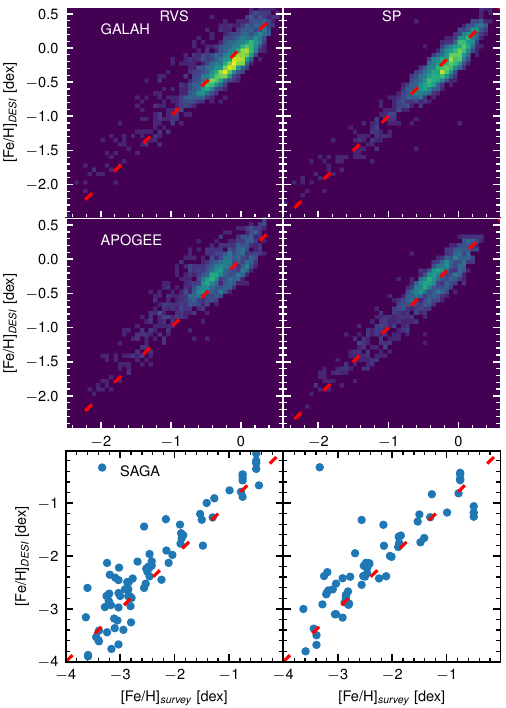}
    \caption{Comparison of iron abundances from DESI's RVS and SP pipelines to the measurements from GALAH (top panels), APOGEE (middle panels) and compilation from SAGA (bottom panels). The left panels show the RVS pipeline measurements, the right panels show the SP pipeline.}
    \label{fig:feh_compar1}
\end{figure}

\begin{figure}
    \centering
    \includegraphics{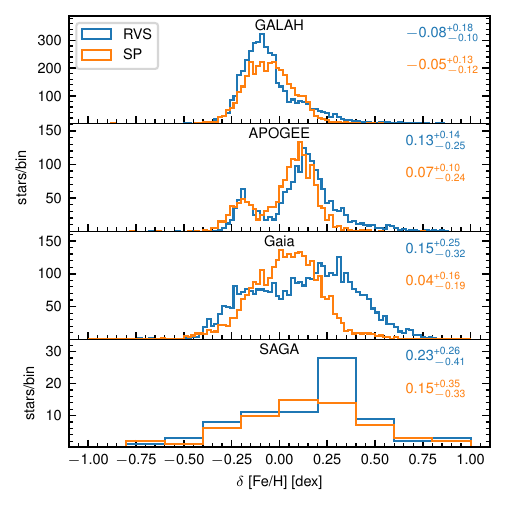}
    \caption{Distribution of differences of uncalibrated [Fe/H] abundances in the VAC with respect to external [Fe/H] measurements.  Each panel shows a histogram of [Fe/H]$_{DESI}$-[Fe/H]$_{external}$ for a different survey. The curves of different colours show either the SP (orange) or RVS (blue) [Fe/H] measurements. The numbers on each panel show the median and the 16-th and 84-th percentiles of the distributions.}
    \label{fig:feh_compar2}
\end{figure}

After investigating the bimodality of the metallicity offsets seen in Figure~\ref{fig:feh_compar1} and \ref{fig:feh_compar2}, we have found that the primary cause is the different metallicity offsets with respect to high-resolution surveys for dwarf and giant stars. The offsets also show some dependence on the effective temperature (see also Figure~9 in \citetalias{Koposov2024_EDR_VAC}). The trends of the metallicity biases with temperature can be well described by a quadratic function of temperature that is different for dwarfs and giants:  
\begin{align}
 {\rm [Fe/H]_{calib}} = {\rm [Fe/H]_{DESI}} - (a + b x + c x^2 ), 
\label{eq:feh_corr}
\end{align}
where $x= 10 \cdot  \log_{10}(T_{\rm eff}\ /5000 {\rm [K]})$ and we adopt different values of $a$, $b$, $c$ for dwarfs and giants.
To find the coefficients, we fitted the stars in DESI that are in common with APOGEE and GALAH (approximately 5000 stars). We optimised the sum of absolute value of deviations of calibrated metallicities with respect to external measurements separately for dwarfs and giants. While doing so, we also varied the $\log g$ value of the transition between the relations for dwarfs and giants. We found the best values of the coefficients for RVS and SP separately. In each case we used the effective temperature and the surface gravity from either RVS or SP tables.  Table~\ref{tab:feh_coeff} gives the best values from this procedure. 


\begin{figure}
    \centering
    \includegraphics{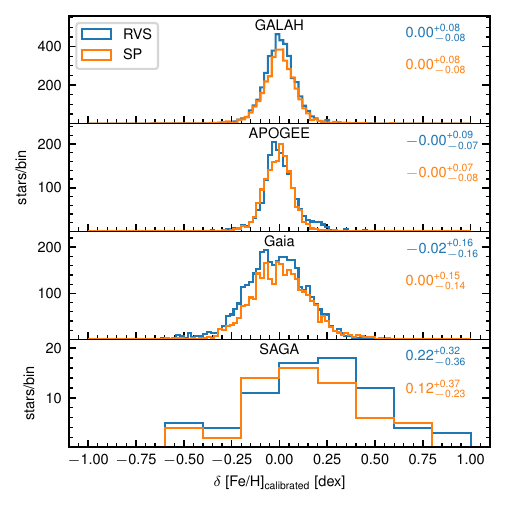}
    \caption{Distribution of differences of calibrated DESI [Fe/H] abundances with respect to external [Fe/H] measurements, after applying a correction based on  Equation~\ref{eq:feh_corr} and Table~\ref{tab:feh_coeff}. Similar to Figure~\ref{fig:feh_compar2} the curves with different colours  blue vs orange show the RVS and the SP pipelines, respectively. The numbers on each panel show the median and the 16-th and 84-th percentiles of the distributions.}
    \label{fig:feh_compar2_calib}
\end{figure}

\begin{table}
    \centering
    \begin{tabular}{ccccc}
    \hline
    Pipeline & Subset & a & b  & c \\
    \hline
        RVS & $\log g <4.3$  & $0.197$ & $-0.809$ & $0.441$ \\
    RVS & $\log g \geq 4.3$ & $0.039$ & $-0.284$ & $0.079$ \\
    SP  & $\log g <3.5$ & $0.113$ & $0.215$ & $-0.146$ \\
    SP  & $\log g \geq 3.5$ & $0.116$ & $-0.216$ & $-0.081$ \\
    \hline
    \end{tabular}

\caption{Metallicity correction coefficients  for Equation~\ref{eq:feh_corr} given for giants and dwarfs and two DESI pipelines. The range of temperatures and surface gravities where these corrections are applicable is given in the text.}
    \label{tab:feh_coeff}
\end{table}

After computing the calibrated [Fe/H] using Eq.~\ref{eq:feh_corr}, we show the distribution of DESI [Fe/H] offsets with respect to other surveys  in Figure~\ref{fig:feh_compar2_calib}.  We see that the bimodality of residuals disappeared and the median offsets for {\it Gaia}, GALAH and APOGEE are now close to zero. The 16/84-percentiles indicate that the residual errors are below 0.1\, dex. The metal-poor SAGA sample, however, still shows an offset of roughly $0.2$\,dex. After calibration of the [Fe/H], the performance of the two pipelines RVS and SP seem very similar based on median offsets and scatter of the distributions.
Given the range of surface gravity and effective temperature of the GALAH and APOGEE stars in common with DESI, we think the corrections that we compute in this section are applicable for stars with $0<\log g<5.1$, $4000<T_{\rm eff}<6600$ for the SP pipeline, and $1.8<\log g<5.5$ and $4200<T_{\rm eff} <6600$ for the RVS pipeline (the parameter ranges refer to the surface gravity and temperature from the RVS and SP measurements, respectively).

The values in the published catalogue are {\bf not} corrected. To make it easier to use  calibrated abundances, we publish a simple Python code {\tt feh\_correct} at \url{https://github.com/desimilkyway/dr1_vac_code/} that performs the corrections described in this section.

\subsection{Radial velocity accuracy and precision}
\label{sec:rv_accuracy}
The quality of DESI radial velocity measurements has already been discussed in \citet{Cooper2023} and \citetalias{Koposov2024_EDR_VAC}. The results of those papers still apply to the survey validation part of the release. What we focus on here is the accuracy of radial velocities for the three programs of the main survey.

Here we discuss several aspects affecting the quality of  DESI radial velocities: 1) possible biases of DESI's radial velocities with respect to other surveys; 2) the typical radial velocity uncertainties for stars in DESI; and 3) the validity of radial velocity uncertainties and the value of the systematic floor in velocity measurements.

Among the three programs of the main DESI survey, 
we highlight the fact that the backup survey is observed in conditions that are drastically worse than the bright survey \citep{backup_paper}. It has been discovered that these conditions (in particular, bright sky with the continuum solar-like spectrum) lead to large biases in the wavelength calibration code of the DESI pipeline (see \citet{guy23,DESI_DR1_2025}). These issues have been identified and rectified in the code of the DESI pipeline, but not in time for DR1. Because of that, the backup program data in DR1 show radial velocity systematics that are substantially worse than other programs, reaching levels of 20\,km\,s${^{-1}}$ in about 1\% of the footprint (see Appendix~\ref{sec:appendix_backup} for more details, and a simplistic position-dependent correction we suggest to use).

\begin{figure}
    \centering
    \includegraphics{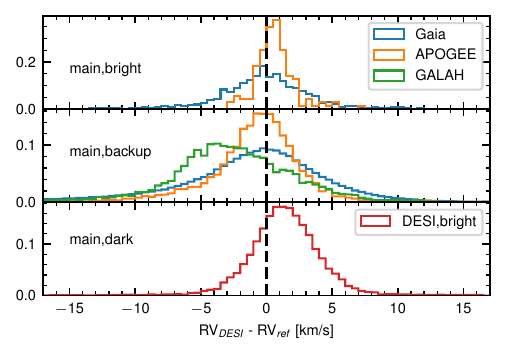}
    \caption{Comparison of the DESI radial velocities from the main survey with respect to external surveys. The panels show the bright (top), backup (middle) and dark (bottom) programs. Since the dark program does not have enough overlap with high-resolution surveys, it is compared to the DESI bright program instead. The backup program is affected by the wavelength calibration systematics described in Section~\ref{sec:validation} with the approximate correction described in Appendix~\ref{sec:appendix_backup}.}
    \label{fig:rv_comp}
\end{figure}

We now compare DESI radial velocities to those from {\it Gaia} RVS, APOGEE, and GALAH. For APOGEE and GALAH we apply the same quality cuts for the reference surveys as described in the previous section. For {\it Gaia} in addition to the previously mentioned cuts, we further require the radial velocity uncertainty to be below 5\,km\,s${^{-1}}$. On the DESI side, we apply the selection criteria listed at the beginning of Section~\ref{sec:validation} and require additionally that the radial velocity uncertainty is below 1\,km\,s${^{-1}}$.
Figure~\ref{fig:rv_comp} shows the distribution of differences of DESI radial velocities with respect to external surveys 
or DESI's own bright program. The bright program was not compared to GALAH due to the small number of overlapping stars -- fewer than 50. Similarly, for the dark program, we compare only with DESI's own measurements because the overlap with GALAH, APOGEE and {\it Gaia} RVS is negligible. 

The top panel of Figure~\ref{fig:rv_comp} shows the velocity offsets for the bright program. From this panel we can see that there is no significant bias in DESI radial velocities, and the observed scatter relative to  APOGEE is $\sim 1$\,km\,s${^{-1}}$. The velocity differences with respect to {\it Gaia} exhibit a negligible offset but a spread of a few km\,s${^{-1}}$ -- a result primarily driven by the larger uncertainties in the {\it Gaia} measurements rather than those from DESI. The median offset for the bright program of the main survey, defined as $V_\mathrm{DESI}-V_\mathrm{external}$, is $-0.19$\,km\,s${^{-1}}$ relative to {\it Gaia}, and $0.41$\,km\,s${^{-1}}$ relative to APOGEE. 

The velocity offsets for the backup program are shown in the middle panel of Figure~\ref{fig:rv_comp}, and the distribution differs significantly from the bright program. Note the very significant tails of the distribution going all the way to 15\,km\,s${^{-1}}$ caused by wavelength calibration issues of the backup program discussed earlier.  While the median offsets for the backup program radial velocities are $-1$, $-0.8$, $-2.8$\,km\,s${^{-1}}$ for {\it Gaia}, APOGEE and GALAH, respectively, they are somewhat misleadingly low as the distribution is asymmetric with a long tail. 
A particularly large median offset with GALAH originates from the spatial structure of the velocity systematics in the backup survey (see Appendix~\ref{sec:appendix_backup}), caused by the fact that different locations on the sky were observed at different times and moon phases.

The bottom panel of the Figure~\ref{fig:rv_comp} shows the distribution of the DESI dark program radial velocity offsets. Since the dark program targets do not have significant overlap with {\it Gaia}, APOGEE, and GALAH, we compare the dark program measurements with those from the bright program ( $\sim 9000$ stars in common). Here we observe a small systematic offset $V_{DESI,dark}-V_{DESI,bright} \sim$ 1.3\,km\,s${^{-1}}$. The exact reason for a velocity offset between dark and bright programs is not clear, but is most likely associated with the wavelength calibration systematics that have been recently corrected, as in the non-public DR2 dataset this offset is close to zero. 

\subsubsection{Radial velocity uncertainties}
\label{sec:rv_prec}
\begin{figure}
    \centering
    \includegraphics{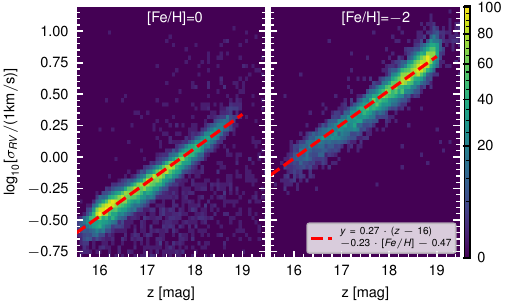}
    \caption{Radial velocity accuracy vs $z-$band magnitude for stars in the bright program of the main survey. The left panel shows metal-rich stars $-0.1<$[Fe/H]$<0.1$, while the right panel shows metal-poor stars $-2.1<$$[Fe/H]$$<-1.9$. We show observations of stars with effective exposure time between 180 and 220 seconds, and effective temperature between 4500 and 7000\,K. The red line on both panels shows the fitted relation to the RV accuracy given in Eqn.~\ref{eq:rv_prec}.}
    \label{fig:rv_prec}
\end{figure}

In \citet{Cooper2023}, we have shown how the radial velocity uncertainty in DESI varies with colour and magnitude for MWS based on survey validation data. We have found, however, that for objects with typical colours $0\lesssim g-r\lesssim 1$, the radial velocity uncertainty mostly depends on the $z$-band magnitude.  The reason for this is that most of DESI's radial velocity information comes from the Z arm of the instrument, due to its higher spectral resolution compared to bluer arms as well as higher total throughput in the red \citep{DESI_Instr2022}.  In Figure~\ref{fig:rv_prec}, we show the radial velocity uncertainty as computed by the RVS vs magnitude for stars in the bright program of the main survey. Here we use the same quality cuts as defined in the beginning of the section, limit the range of effective temperatures to $4500$\,K$<T_{\rm eff}<7000$\,K (roughly corresponding to the colour range of $0\lesssim g-r\lesssim 1.1$), and restrict the sample to those stars with effective exposure times close to the typical one of the bright survey (i.e. between 180 and 220 sec).\footnote{The effective exposure time for every source can be computed as $12.15 \cdot$ {\tt TSNR2\_LRG} \cite[see][for more details]{guy23}, where the {\tt TSNR2\_LRG} column is available in the {\tt SCORES} extension of the catalogue.} Since the radial velocity precision should be a function of metallicity, we split the plot between high metallicity $-0.1<{\rm [Fe/H]}<0.1$\,dex (left panel) and low metallicity $-2.1<{\rm [Fe/H]}<-1.9$\,dex stars (right panel). As expected, high metallicity stars provide more precise radial velocities at the same magnitude due to a higher amount of spectral information caused by larger number and stronger  absorption lines.

In the Figure we also overplot the approximate relation for the radial velocity uncertainty that incorporates the radial velocity precision change with metallicity. This relation was fitted to all the DESI stars in the bright program with  $4500<T_{\rm eff}/K<7000 $:
\begin{align}
    \log_{10} \frac{\sigma_{RV}}{1\, \mathrm{km\,s^{-1}}} = -0.47 + 0.27\ (z-16) - 0.23\ {\rm [Fe/H]}  
\label{eq:rv_prec}
\end{align}

We remark that the slope of the radial velocity uncertainty as a function of magnitude is slightly greater than 0.2. For background-dominated spectra, a slope of 0.4 is expected, while signal-dominated cases should yield a slope of 0.2. Thus the bright program is as expected closer to  the signal-dominated case. The relation defined in Eq.~\ref{eq:rv_prec} has been derived for the typical effective exposure time of the bright program, but there are sources in the catalogue with larger effective exposure times (for example if they have been observed multiple times over several passes). In this case the uncertainties need to be scaled from Eq.~\ref{eq:rv_prec} by $\sqrt{{\rm t_e}/200}$  where $t_e$ is effective exposure time for the source in seconds.

While Figure~\ref{fig:rv_prec} provides the analysis for the bright program, we do not show the corresponding plot for the backup program as its observing conditions are far less predictable, and the radial velocities in that program are subject to the systematic issues discussed earlier.

We do not present a similar figure for the uncertainties of the [Fe/H] abundances since they are more affected by systematic errors. However, we find that the formal [Fe/H] uncertainty (for the RVS pipeline) can be similarly well described by a relation similar to Eqn~\ref{eq:rv_prec}: 
\begin{align}
    \log_{10} {\sigma_\mathrm{[Fe/H]}} = -1.83 + 0.27\ (z-16) - 0.23\ {\rm [Fe/H]}  
\label{eq:feh_prec}
\end{align}

We notice that, remarkably, the magnitude and metallicity dependence slopes are identical to the ones determined for the radial velocity uncertainty.

\subsubsection{Systematic floor of radial velocity measurements}

Another important consideration for  understanding the quality of DESI's radial velocities is whether the reported radial velocity uncertainties accurately reflect true measurement uncertainties and whether any additional instrument systematic effects contribute to the radial velocity errors. One can approach this by analysing repeated radial velocity measurements of the same objects. 
Since in this VAC we include, for the first time, the measurements from single-epoch observations, we are now able to perform this analysis. Crucially, this type of test bypasses questions about absolute radial velocity offsets relative to ground truth (and whether such offsets vary from star to star), focusing instead solely on the repeatability of radial velocities and their associated uncertainties.

A common approach to characterizing the repeatability is to assume that the calibrated radial velocity uncertainties $\sigma_\mathrm{calib}$ can be expressed as a function of the formal uncertainty $\sigma$ through the following expression: 

\begin{align}
\sigma_{\rm calib} = \sqrt{ (a \,\sigma)^2 + f^2}\label{eqn:vel_calib},    
\end{align}
where the factor $a$ scales the uncertainty and $f$ refers to a systematic floor \citep[see, e.g.,][]{Simon2007,Li2019}. If the data reduction pipeline correctly propagates all the uncertainties and the radial velocity measurement correctly takes them into account, then one expects $a=1$. The systematic floor can be interpreted as an irreducible Gaussian noise ${\mathcal N}(0,f)$ added to every radial velocity measurement. The most common causes for that floor on various instruments are inaccuracies in the wavelength calibration or flexures of the spectrograph.

It is possible to determine the constants $a$ and $f$ if we consider all pairs  $(i,j)$ of observations for each star with radial velocities $v_i,v_j$ and uncertainties $\sigma_i,\sigma_j$. Assuming a Gaussian distribution, and calibration of uncertainties defined in Eqn.~\ref{eqn:vel_calib}, one can then compute that 
  $\frac{1}{\sqrt{2}} (v_i-v_j) $ should be normally distributed following  ${\mathcal N} \left(0, \sqrt{a^2 \frac{\sigma_i^2+\sigma_j^2}{2} +  f^2}\right)$.

\begin{figure}
    \centering
    \includegraphics{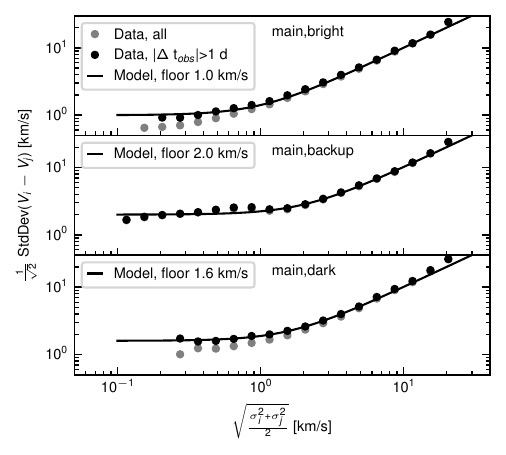}
    \caption{Radial velocity accuracy based on repeated observations in DESI. The black points show the estimate of the standard deviation of pair-wise velocity differences (divided by $\sqrt2$) in bins of the combined radial velocity uncertainty $\frac{1}{\sqrt{2}}\sqrt{\sigma_1^2+\sigma_2^2}$. Different panels (from top to bottom) show bright, backup and dark programs, respectively. Grey points show measurements from all repeats, while dark points show measurements from repeated observations that are separated by more than one day. The measurement of standard deviation is done through 84-th and 16-th percentiles of the distribution. Black curves show the error model of $\sigma_{\rm calib} = \sqrt{x^2 + f^2}$ with different floor values $f$ (1.0, 2.0, 1.6\,km\,s$^{-1}$) for bright, backup, and dark programs, respectively. }
    \label{fig:repeats}
\end{figure}

Motivated by this, in Figure~\ref{fig:repeats}  we show the robust estimate of the standard deviation of pair-wise radial velocity differences $\frac{1}{\sqrt{2}} {\rm StdDev}[v_i-v_j]$ in bins of $ \sqrt{\frac{\sigma_i^2+\sigma_j^2}{2}}$ \citep[see i.e.][for a similar approach]{koposov2011}. In the case of perfect uncertainties and no systematic floor the points should lie along the y=x line.

To ensure robustness against outliers (e.g. such as those introduced by binaries), the standard deviation of pairwise differences is estimated as 
half of the difference between 84-th and 16-th percentiles of the distribution. Grey points show measurements using all pairs of observations within the corresponding bin of $ \sqrt{\frac{\sigma_i^2+\sigma_j^2}{2}}$, while black points include only pairs of observations separated by more than a day. 
The top, middle, and bottom panels show the results from bright, backup, and dark programs, respectively.  
As expected, the measured standard deviations flatten at small $\sqrt{\frac{\sigma_i^2+\sigma_j^2}{2}}$ uncertainties (left side of the panels), indicating the presence of a systematic floor in radial velocity measurements. Additionally, the black points tend to lie above the grey points, as they include only pairs of observations separated by more than a day. This is expected if certain systematic errors are specific to a given night -- for example, due to the particular arc calibrations used on that night.
To determine the parameters $a$ and $f$ for different programs, we fit Eqn.~\ref{eqn:vel_calib} to the black data points shown in the Figure.\footnote{The reason to fit for the systematic floor from black points that are above grey points is that we think the measurements of velocity differences from different nights provide a more honest assessment of systematics} In all cases, we have found that the best-fit value of the parameter $a$ (responsible for scaling of the uncertainties) is within a few percent of 1, hence we adopt $a=1$. The values of the systematic floor $f$ measured for different programs are given in each panel of the Figure. We also show the best-fit models $\sqrt{x^2 +f ^2}$ for each program on the Figure, demonstrating that the Eqn.~\ref{eqn:vel_calib} can well describe the observed behaviour.

The systematic floor for the bright program is $1.0$\,km\,s$^{-1}$ (note that if we consider the floor measurement from all observations including observations on the same night the floor is $0.65$\,km\,s$^{-1}$, significantly better than 1\,\kms).  
The floor for the backup program is around 2\,km\,s$^{-1}$; however, this value is somewhat misleading. As shown in Figure~\ref{fig:rv_comp} and Appendix~\ref{sec:appendix_backup}, the backup program exhibits non-Gaussian systematic errors, making the formalism defined in Eqn.~\ref{eqn:vel_calib} inappropriate for characterizing its uncertainties.
For the dark program the floor is found to be $1.6$\,km\,s$^{-1}$ which is higher than a systematic floor of $\sim 1$\,km\,s$^{-1}$ for the bright program. The reason for this is not yet understood.

\begin{figure}
    \centering
    \includegraphics{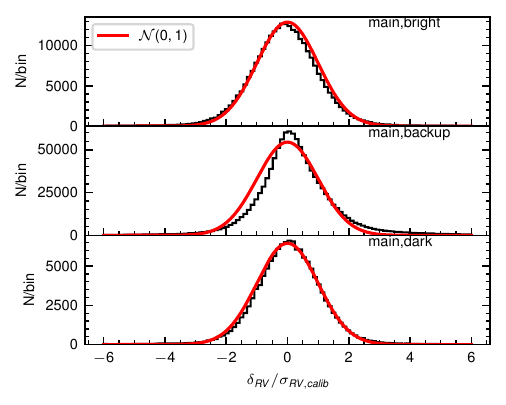}
    \caption{The distribution of pairwise velocity differences divided by the combined calibrated uncertainty $\frac{v_i -v_j}{\sqrt{\sigma_{i,{\rm calib}}^2+\sigma_{j,{\rm calib}}^2}}$ in different programs of the main survey. Here we only show observations separated by more than a day. The distributions are slightly asymmetric, because we only consider differences between pairs $v_i-v_j$ where i-th observation is before j-th. The red curve shows the expected ${\mathcal N}(0,1)$ distribution.}
    \label{fig:repeat_delta}
\end{figure}

As a last check of the calibrated radial velocity uncertainties, we look at the distribution of pair-wise velocity differences divided by the combined calibrated uncertainty $\frac{v_i-v_j}{\sqrt{\sigma_{{\rm calib},i}^2+\sigma_{{\rm calib},j}^2}}$. We consider only observations separated by more than one day. If the calibrated uncertainties are accurate, the distribution should be close to normal with a mean of zero and unit variance. This is shown in Figure~\ref{fig:repeat_delta} for three programs of the main survey. For both bright and dark programs, the distributions are close to a Gaussian, confirming that our calibration procedure is accurate. For the backup program, however, the distribution is clearly non-Gaussian, but this is expected due to the systematic issues discussed earlier.

Finally, we note that the radial velocity uncertainties reported in the VAC do not include the correction for the systematic floor, and users of the catalogue should apply it if it is appropriate for their science case.

\section{Discussion}
\label{sec:discussion}
\subsection{Science with the DR1 stellar sample}
In this paper, we presented the stellar catalogue constructed based on the analysis of stellar spectra in DESI DR1 with stellar parameters, abundances, and radial velocity measurements. For the first time, the catalogue reports single-epoch measurements.

While not discussed in this paper, the DR1 VAC is also accompanied by several additional value-added catalogues that can be useful for the interpretation of the DESI data. Specifically: 

\begin{itemize}
    \item Spectrophotometric distances for DESI stars. They are available in two independently derived VACs (SpecDis and SPDist). One is based on the neural network trained on the DESI spectra and {\it Gaia} parallaxes \citet{Li2025_Specdist}, and another trained on the SP stellar parameters, calibrated on the globular clusters and dwarf galaxies \citep{Thomas2022}. Both methods are able to provide distances for a large number of stars with better than 25\% {precision}.
    \item The catalogue of $\sim$ 10,000 blue horizontal branch stars that are spectroscopically confirmed in DR1, based on the methodology presented in \citet{bystrom24a}. 
\item The catalogue of radial velocities and metallicities of more than 6,000 RR Lyrae stars in DESI based on multi-epoch observations \citep{Medina2025}.
\end{itemize}

The samples of BHBs, RRLs and halo stars with spectroscopic distance estimates provide a useful set of stars for chemo-dynamic studies of the MW halo, substructure search, and constraining the MW potential and Dark Matter distribution \citep[see, e.g.,][]{Naidu2020,dodd2023,xue2008}. 
Due to DESI's depth, the BHB and RRL catalogs probe distances beyond 100 kpc \citep{bystrom24a,Medina2025}. Similarly, samples of red giant branch stars in DESI probe a large range of Galactocentric distances. If we use the distances from the VAC by \citet{Li2025_Specdist} and require better than 30\% distance {precision} we can estimate that MWS has more than  50,000 stars with heliocentric distances larger than 10\,kpc, and more than 3,000 stars with distances larger than 50\,kpc. 

The large number of metallicities measured by DESI for halo stars enables studies of the metallicity distribution functions of different stellar halo components and their changes throughout the Galaxy. Also, as shown in Figure~\ref{fig:feh_distr}, DESI detects a large number of  metal-poor stars ($\sim $ 110,000 stars with [Fe/H]$<-2$ and  $\sim$ 5,000 stars with [Fe/H]$<-3$). This enables not only the search for some of the most metal-poor stars in the Milky Way \citep{Caffau2011}, but also exploration of the properties of much larger homogeneous samples of extremely metal-poor stars than previously available \citep{aguado2019,youakim2020}.

Aside from studying the overall stellar halo, the data from this VAC already contains a large number of stars that are members of various substructures in the Galaxy such as globular and open clusters, dwarf galaxies, and stellar streams. To illustrate that, in Table~\ref{tab:members} we list the systems with more than 10 members or likely members that have been observed by DESI and are included in this VAC. We used the catalogues from \citet{CantatGaudin2020}, \citet{Baumgardt2021},  \citet{Ibata2024} and \citet{PaceLi2022} for open clusters, globular clusters, stellar streams and dwarf galaxies, respectively. When the membership probability $P$ was available, we only considered stars with $P>0.9$.  We can see that this VAC contains more than 10 likely members for at least 16  clusters, 24 streams, and 4 dwarf galaxies.\footnote{There are also several ultra-faint dwarf galaxies with $1-5$ potential members, such as Bootes 2, Bootes 3, Hercules 1, Leo 4, Leo 5, Pegasus 3, Ursa Major 1, and Willman 1.}  Particularly notable is the large number of potential stellar stream members in DESI.\footnote{We expect this number is underestimated due the incompleteness of the catalogue of \citet{Ibata2024} and new streams being discovered, such as \citet{tian2024}.} The key reason for this is the magnitude range of the bright survey $16<r\lesssim 19-20$ and high completeness of DESI in that magnitude range. 
We also note that the number of open clusters with DESI data is severely underestimated, because the open cluster member catalogue \citet{CantatGaudin2020} is incomplete at faint ($G>17-18$) magnitudes \citep{Hunt2023}.

\begin{figure}
    \centering
    \includegraphics{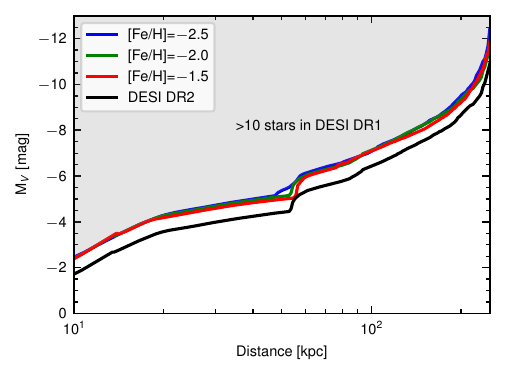}
    \caption{The absolute magnitude limit for a stellar structure at different heliocentric distances to have more than 10 stars in DESI DR1. The curves with different colours show limits for different [Fe/H]. The calculation relies on the DESI DR1 completeness of 20\% within the currently observed footprint above $|b|=40^\circ$ for stars with $16<r<19$ that are further than 10\,kpc. The black curve shows the expected magnitude limit for DESI DR2 which will have completeness of $\sim 40\%$. The magnitude limit shown on the figure applies irrespective of whether the structure is a dwarf galaxy, a stream, and how dispersed it is within the footprint .}
    \label{fig:maglim}
\end{figure}

To illustrate  DESI's efficiency in sampling streams and galaxies in the MW halo, in Figure~\ref{fig:maglim} we show what would be the required absolute magnitude of a stellar system within the DESI footprint at different heliocentric distances to have more than 10 stars in the DESI catalogue. To calculate this we simulate 10\,Gyr old stellar populations with different metallicities using MIST isochrones \citep{Choi2016,Dotter2016} and Chabrier initial mass function \citep{chabrier2003}. We place the systems at different heliocentric distances and then use the DESI DR1  spectroscopic targeting completeness of $\sim 20\%$ for main sequence and distant red-giant branch stars with  $16<r<19$ at high galactic latitude ($|b|>40^\circ$).\footnote{The only case where the assumption of constant spectroscopic completeness would break down is the case of an extremely dense stellar system such as a dense star cluster.} We assume the systems are within the {currently observed} DR1 footprint above a galactic latitude of 40 degrees. Curves of different colours show magnitude limits for different metallicities of [Fe/H]$=-1.5$, $-2$ and $-2.5$ for the stellar populations. We also show the curve expected for DR2 where the spectroscopic completeness for $16<r<19$ targets within an even larger footprint on the sky is close to 40\%.  The change of the magnitude limit with distance is driven by changes in the number of stars in the DESI magnitude range and the feature at 50\, kpc is due to the loss of the main sequence turn-off stars.
We can see that with existing DESI sampling we expect to detect more than 10 stars in streams or dwarf galaxies that are brighter than $M_V\sim -4$ (within the footprint) at 20 kpc, and brighter than $M_V\sim -8$ at 100\, kpc. This magnitude limit will shift by approximately one magnitude for the DR2. This shows that DESI gives us a uniquely rich view of all the accreted systems in the MW halo, such as streams and faint dwarfs. We note, however, that while in our calculation we only check which systems will have more than 10 stars observed by DESI, not all of these substructures may be easily separated from the background.

While Table~\ref{tab:members} gives just the number of stars in various clusters, dwarf galaxies, and stellar streams in DESI, each member star has a DESI metallicity and radial velocity measurement. The detailed analysis of these is beyond the scope of the paper, but in the Appendix~\ref{sec:appendix_fehs}, we measure the mean abundances and intrinsic metallicity spread in each system using a mixture model and calibrated abundances from DESI.  Table~\ref{tab:members_full} provides the mean metallicity of each system and intrinsic metallicity spread for both SP and RVS pipelines. A more detailed analysis is coming in future papers.

\begin{figure*}
    \centering
    \includegraphics[]{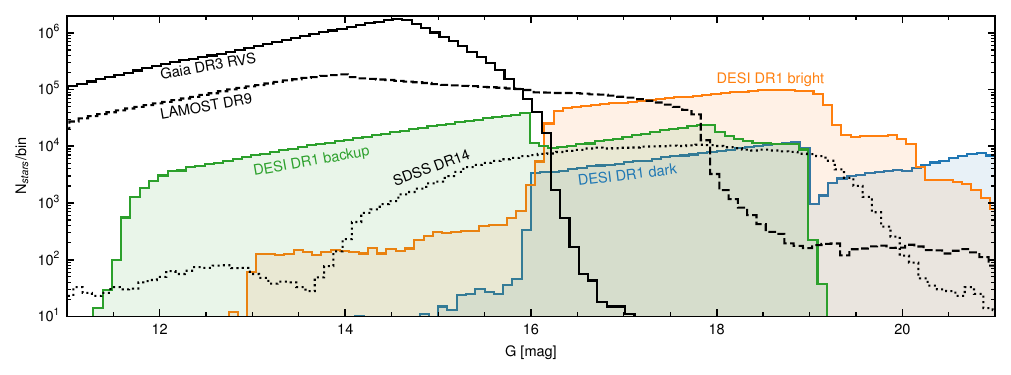}
    \caption{The magnitude distribution of stars with radial velocity and abundances from existing low and medium-resolution surveys. We show the DESI DR1 data from the main survey included in this VAC by blue, green, orange lines for the backup, dark and bright program respectively. The distribution for the DR2 (not shown on the figure) is expected to be $\sim$ 3 times larger than that for DR1. For SDSS (shown with a dotted line) we use the DR14 \citep{apogee2018} objects that have radial velocity measurements in the {\tt sppparams} table. For LAMOST DR9 (dashed line), we use objects in the stellar catalogue of the low-resolution survey. For {\it Gaia}  RVS (solid black line) we use all objects with the radial velocity measurement present in {\it Gaia} DR3 catalogue.}
    \label{fig:survey_comp}
\end{figure*}

In this paper, we mostly focused on measurements from coadded spectra, but as discussed in Section~\ref{sec:content}, the current catalogue contains more than a million stars with repeated radial velocity and stellar parameter measurements. While the backup program that contains many of these repeats suffers from larger systematic radial velocity errors, the observations from the bright and dark survey are already sufficient for radial velocity variability studies, such as investigation of the binary population \cite[e.g., similar to ][]{Badenes2018} as a function of metallicity and environment, searches for rare systems with massive companions \citep{ElBadry2023} and many others. A particular example of a study relying on the multi-epoch nature of DESI data are the RR Lyrae analyses published in \citet{Medina2025,Medina2025b}.

\begin{table*}
\centering
\caption{The number of stars from various objects in the DESI dataset. We use the list of potential members of dwarf galaxies from \citet{PaceLi2022}, potential globular cluster members from \citet{Baumgardt2021}, open cluster members from \citet{CantatGaudin2020}, and candidate members of stellar streams from \citet{Ibata2024}. For dwarf galaxy and cluster members, a probability threshold larger than 0.9 was applied.}
\begin{tabular}{lll|lll|lll}
\hline
\textbf{Object Type} & \textbf{Name} & \textbf{Number} &
\textbf{Object Type} & \textbf{Name} & \textbf{Number} &
\textbf{Object Type} & \textbf{Name} & \textbf{Number} \\
\hline

dSph & Canes Venatici 1 & 16 & GC & M 2 & 151 & Stream & Gaia-7 & 18 \\
dSph & Draco 1 & 180 & GC & Pal 5 & 50 & Stream & Fjorm & 35 \\
\cline{4-6}
dSph & Sextans 1 & 367 & OCL & NGC 2632 & 52 & Stream & Orphan & 42 \\
dSph & Ursa Major 2 & 12 & OCL & Melotte 111 & 15 & Stream & Gaia-1 & 32 \\
\cline{1-3}

GC & NGC 2419 & 32 & OCL & NGC 2539 & 42 & Stream & LMS-1 & 41 \\
\cline{4-6}
GC & M 53 & 131 & Stream & Kwando & 11 & Stream & GD-1 & 280 \\
GC & NGC 5053 & 58 & Stream & Leiptr & 18 & Stream & Gaia-6 & 52 \\
GC & M 3 & 141 & Stream & C-25 & 19 & Stream & New-19 & 12 \\
GC & NGC 5466 & 46 & Stream & C-10 & 40 & Stream & Pal 5 & 24 \\
GC & NGC 5634 & 39 & Stream & Gjoll & 14 & Stream & Svol & 23 \\
GC & M 5 & 853 & Stream & C-24 & 36 & Stream & M 92 & 28 \\
GC & M 13 & 623 & Stream & Slidr & 48 & Stream & Hrid & 23 \\
GC & M 12 & 28 & Stream & Ylgr & 23 & Stream & New-21 & 45 \\
GC & M 92 & 202 & Stream & Sylgr & 105 & Stream & Phlegethon & 38 \\
GC & M 15 & 77 & Stream & New-15 & 14 & & &  \\

\hline
\end{tabular}
\label{tab:members} \end{table*}

The DESI data have been already demonstrated to be extremely valuable for the analysis of stellar streams \citep{Valluri2025, aganze2025}, detection of substructures in the MW halo \citep{Kim2025}, studies of the large scale kinematics of the Galactic halo \citep{bystrom24a}, studies of white dwarfs \citep {Manser2024}, studies of the stellar halo in M31 \citep{Dey2023}, and detection of very metal-poor stars \citep{AllendePrieto2023}. Upcoming publications based on DR1 and DR2 data include studies of the dwarf spheroidal galaxies in DESI DR1, studies of the RRL population, and studies of the perturbations of the MW disk. 

\subsection{Comparison with other surveys}
\label{section:comparison_other_surveys}
We now place the DESI MWS and this VAC in the context of other recent and ongoing spectroscopic surveys with publicly available data. Since DESI spectroscopic resolution varies from 2000 in the blue to 5000 in the red, DESI should be compared to low- and medium-resolution surveys rather than high-resolution surveys like Gaia-ESO \citep{gilmore2012}, GALAH \citep{galah0},  APOGEE \citep{majewski17a}.  
Three major surveys that publicly released data with comparable resolution are LAMOST with its low and medium-resolution mode \citep{Lamost2012},   
{\it Gaia} RVS \citep{katz2023} and SDSS/SEGUE \citep{yanny2009, rockosi2022}. In addition, there are also several surveys that have either just started or are about to start such as SDSS-V  with Milky Way Mapper \citep{kollmeier2017}, WEAVE \citep{jin2023}, and 4MOST \citep{dejong2019}. All of these have low- as well as high-resolution modes. Their low-resolution programs are expected to cover a similar magnitude range as DESI.  

The key advantage of DESI, in comparison to other existing datasets, is that DESI (and its bright program) is significantly deeper and has significantly more faint stars in $17<G<21$ than existing surveys. Figure~\ref{fig:survey_comp} shows the magnitude distribution for stars in DESI DR1 (separated by program) and compares it to other large low and medium resolution surveys: LAMOST, {\it Gaia} and SDSS/SEGUE. The Figure clearly shows an increase by an order of magnitude in the number of fainter stars provided by DESI compared to existing samples. We also remark that while the number of stars at $16<G<17.5$ is comparable between LAMOST and DESI (DR1), DESI has a factor of 10 better radial velocity accuracy at these magnitudes $<1$\,km\,s$^{-1}$ vs 5-10\,km\,s$^{-1}$ in LAMOST \citep{Xiang2015,tsantaki2022}.

For brighter stars with $G<16$, while {\it Gaia} RVS have typically accurate radial velocities comparable to DESI and it has an order of magnitude more stars  than DESI ($\sim$ 33 million vs $\sim$ 1 million), the backup program of DESI provides a much wider wavelength coverage from 3600\,\AA\ to 9800\,\AA\ compared to 8500\AA-8700\AA\ for {\it Gaia} RVS.

While currently DESI dominates in terms of the number of faint stars with spectroscopy, the upcoming surveys like WEAVE and 4MOST will observe a large number of stars in similar magnitude range to DESI and will have somewhat higher resolution. The number of stars delivered by these surveys is expected to be a few million for WEAVE and $\sim 15 $ million for 4MOST. 
\subsection{Future}
\label{sec:future}

{As of April} 2025, the DESI survey is still ongoing -- in its fourth year of operation of the main survey. With the bright program of the main survey ahead of schedule and having finished the planned four passes of the survey, it is currently performing the fifth pass of the sky. The backup program is currently 70\% done. The next public data release from DESI will be based on data accumulated over three years of the DESI survey (observed before Summer 2024). This release is already available to the collaboration and is under internal validation. It is expected to be released in about 2 years (Spring 2027). This release will have more than 12 million stars across all the programs and will have significant processing improvements. 

\section{Conclusion}

We present the stellar component of the first data release of DESI, and the Value-Added Catalogue with radial velocity, abundance and stellar parameter measurements for more than four million stars. The sample of measured stars spans a large range of magnitudes, $12\lesssim G\lesssim 19$ for the backup program of the main survey ($\sim $ 1.2 million stars), and $16\lesssim G\lesssim 21$ for the bright and dark program of the main survey ($\sim$ 3 million stars).

\begin{itemize}
\item This value-added catalogue presents an order of magnitude increase in the number of faint stars $17.5<G<21$ with radial velocity and abundance measurements compared to existing spectroscopic surveys.
\item More than a million stars in the catalogue have repeated measurements of radial velocities and stellar parameters from the RVS pipeline over the timespan of a year, enabling variability studies.
\item The bright program of the DESI main survey provides highly accurate radial velocities with formal uncertainties ranging from 0.1 to 6\,km\,s$^{-1}$ (median uncertainty of 0.6\,km\,s$^{-1}$). The systematic floor of the radial velocities was measured from repeated observations to be $\sim 1$\,km\,s$^{-1}$. 
\item The uncalibrated DESI [Fe/H] measurements in the catalogue from two pipelines show typical accuracy of $0.15-0.2$\,dex when comparing to high-resolution surveys. After temperature and surface-gravity dependent calibration, DESI metallicities show agreement with high-resolution surveys that is better than 0.1\,dex, albeit with somewhat larger systematic errors for metal-poor stars.
\item This catalogue contains the first release of measurements of 1.2 million stars observed in the Milky Way Backup Program \citep{backup_paper}. This is the survey of mostly brighter stars $12<G\lesssim 16$ extending to $G= 19$, done in bad observing conditions. The velocity measurements for these stars  in the current release are affected by the wavelength calibration issues that will be rectified in the next release. In the paper, we provide an approximate correction for these velocity systematics.
\item This VAC contains a large number of metal-poor stars, with more than $\sim $ 100,000 stars having [Fe/H]$<-2$ and several thousands of stars with [Fe/H]$<-3$.

\item  With the current DESI spectroscopic completeness of $\sim$ 20\% within the observed footprint for distant $16<r<19$ stars, we expect DESI to have observed more than 10 stars in every system such as globular cluster, dwarf galaxy, or stream that is within the footprint and brighter than $M_V<-4$ at 20\,kpc and brighter than $M_V<-6$ at 60\,kpc. These numbers will improve by two magnitudes for DESI DR2. 
\item The power of DESI in detecting faint substructures in the MW halo is clear from the fact that even this data release contains stars that are likely members of at least several dozen stellar streams (each of them with more than 10 stars). This VAC also contains thousands of stars from other known substructures in the Milky Way, including 4 dwarf galaxies, 13 globular clusters and several open clusters and is perfectly suited for galactic archaeology studies. We provide the measurement of the mean metallicities and intrinsic metallicity spread of each system based on published astrometric member catalogues.

\item The stars in the catalogue sample many major galactic components, such as thin disc, thick disc and stellar halo. Due to the magnitude range of stars in the bright program, DESI has observed a large number of tracers in the distant stellar halo. Thousands of RRL and BHB stars have been observed in DESI and extend all the way beyond 100\,kpc.  Main sequence and red giant branch samples in DESI have more than 50,000 stars farther than 10\,kpc, and several thousands farther than 50\,kpc.
\item The list of known limitations and data quality issues in the  released catalogue and associated stellar spectra is documented in the  Appendix~\ref{sec:known_issues}.
\item The next public data release of DESI is planned for 2027 and will have three times as many stars as the DR1.
\end{itemize}

\section*{Acknowledgements}

SK acknowledges support from the Science \& Technology Facilities Council (STFC) grant ST/Y001001/1. 
CAP acknowledges financial support from the Spanish MICIU projects PID2020-117493GB-I00 and PID2023-149982NB-I00. 
T.S.L. acknowledges financial support from Natural Sciences and Engineering Research Council of Canada (NSERC) through grant RGPIN-2022-04794. LBeS acknowledges support from CNPq (Brazil) through a research productivity fellowship, grant no. [304873/2025-0].

This paper made use of the Whole Sky Database (wsdb) created and maintained by Sergey Koposov at the Institute of Astronomy, Cambridge with financial support from the Science \& Technology Facilities Council (STFC) and the European Research Council (ERC). 

This material is based upon work supported by the U.S. Department of Energy (DOE), Office of Science, Office of High-Energy Physics, under Contract No. DE–AC02–05CH11231, and by the National Energy Research Scientific Computing Center, a DOE Office of Science User Facility under the same contract. Additional support for DESI was provided by the U.S. National Science Foundation (NSF), Division of Astronomical Sciences under Contract No. AST-0950945 to the NSF’s National Optical-Infrared Astronomy Research Laboratory; the Science and Technology Facilities Council of the United Kingdom; the Gordon and Betty Moore Foundation; the Heising-Simons Foundation; the French Alternative Energies and Atomic Energy Commission (CEA); the National Council of Humanities, Science and Technology of Mexico (CONAHCYT); the Ministry of Science, Innovation and Universities of Spain (MICIU/AEI/10.13039/501100011033), and by the DESI Member Institutions: \url{https://www.desi.lbl.gov/collaborating-institutions}. Any opinions, findings, and conclusions or recommendations expressed in this material are those of the author(s) and do not necessarily reflect the views of the U. S. National Science Foundation, the U. S. Department of Energy, or any of the listed funding agencies.

The authors are honored to be permitted to conduct scientific research on I'oligam Du'ag (Kitt Peak), a mountain with particular significance to the Tohono O’odham Nation.

{This work has made use of data from the European Space Agency (ESA) mission
{\it Gaia} (\url{https://www.cosmos.esa.int/gaia}), processed by the {\it Gaia}
Data Processing and Analysis Consortium (DPAC,
\url{https://www.cosmos.esa.int/web/gaia/dpac/consortium}). Funding for the DPAC
has been provided by national institutions, in particular the institutions
participating in the {\it Gaia} Multilateral Agreement.}

\section*{Data Availability}

The value-added catalogue described in the paper is available at \url{https://data.desi.lbl.gov/doc/releases/dr1/vac/mws/}.
All the plots for this paper can be reproduced using code available at \url{
https://github.com/desimilkyway/dr1_vac_plots}, with the data needed for reproducing the plots available at \url{https://doi.org/10.5281/zenodo.15317972}.
The notebook with the tutorial on using the DR1 VAC is available at \url{https://github.com/desimilkyway/dr1_tutorials}. The radial velocity corrections for the backup program observations are available at \url{https://doi.org/10.5281/zenodo.15469272}.

\section*{Software}
{\tt astropy} \citep{astropy:2013, astropy:2018, astropy:2022}, 
{\tt duckdb} \citep{duckdb}, 
{\tt dynesty} \citep{dynesty1,dynesty2}, 
{\tt ipython} \citep{ipython}, 
{\tt h5py} \citep{h5py}, 
{\tt healpy} \citep{healpy}, 
{\tt matplotlib} \citep{matplotlib},
{\tt minimint} \citep{minimint}, 
{\tt numdifftools} \url{https://numdifftools.readthedocs.io/en/master/}, 
{\tt numpy} \citep{numpy}, 
{\tt imf} \url{https://github.com/keflavich/imf}, 
{\tt pytorch} \citep{pytorch2019}, 
{\tt rvspecfit} \citep{rvspecfit}, 
{\tt scipy}  \citep{scipy}, 
{\tt sqlutilpy} \citep{sqlutilpy},
{\tt q3c} \citep{q3c}
\bibliographystyle{mnras}

\bibliography{bibliography}

\appendix
\section{Main catalogue file}

\label{sec:appendix_mwsall}

The file {\tt mwsall-pix-iron.fits} is the main catalogue of measurements from coadded DESI spectra containing measurements from all surveys and programs.  This FITS file has several extensions. They are described in detail in \citetalias{Koposov2024_EDR_VAC}, but we repeat a brief summary here. The {\tt RVTAB} extension contains the measurements from the RVS pipeline, such as radial velocities, stellar parameters, key target information such coordinates, primary flag, the warning flag from the RVS pipeline, the survey and program the star is coming from. The {\tt SPTAB} extension contains the measurements of stellar parameters and individual elements by the SP pipeline.
The {\tt FIBERMAP} extension contains the targeting information, such as coordinates, targeting bitmasks such as {\tt MWS\_TARGET}, Legacy survey photometry,  some {\it Gaia} information used for targeting  (partially based on {\it Gaia} DR2 and partially on DR3). The extension also contains information about the number of coadded exposures and total exposure time.
The {\tt GAIA} extension contains the crossmatch with {\it Gaia} DR3 with all the columns from the {\tt gaia\_source} table.
The {\tt SCORES} extension contain various quality assurance quantities on spectra, exposures as well as  information necessary to compute effective exposure times.
Each extension has the same number of rows (6,372,607). Each extension is row matched to another, such that i-th row in every extension refers to the same spectrum.

\section{Known issues}

\label{sec:known_issues}
Here we list the known issues that are relevant for the users of this VAC and stellar spectra. 
\begin{itemize}

    \item The resolution matrices \citep[see][]{guy23} for the coadded spectra were calculated incorrectly by the DESI pipeline. This may lead to biases in resolution matrices, in particular, next to masked pixels and in the areas of highly variable signal-to-noise ratio. This will lead to reduced quality of model spectra if the resolution matrix is used in the forward modelling.
    \item Due to a bug in the coaddition code of the DESI pipeline, when high signal-to-noise spectra with sharp absorption lines are coadded, some pixels around the line may be masked as cosmic rays. That may affect line profile shapes.
    \item Due to a bug in the flux-calibration part of the pipeline, the variances of all flux-calibrated spectra are underestimated. The effect is negligible at low signal-to-noise, but is becoming important for S/N above 25-50. That means that most DESI spectra/pixels that have formal S/N larger than 50 have actual S/N  $\lesssim 50$.
\item While the {\tt Redrock} spectroscopic classification is very robust and $\tt  RR\_SPECTYPE=STAR$ provides a very pure sample of stars, for faint sources with signal-to-noise below 5, {\tt Redrock } may misclassify up to 10-20\% of stars as non-stars.
    \item The radial velocities of stars in the backup program are affected by wavelength calibration systematic biases (see Appendix~\ref{sec:appendix_backup}).  We provide the lookup table to approximately correct these biases.
    \item The [Fe/H] from both pipelines shows $\log g$, $T_\mathrm{eff}$ dependent systematics when comparing to external surveys. We provide the simple corrections that reduce the systematic errors below 0.1 dex.
    \item $\mathrm{[\alpha/Fe]}$ abundances at low metallicities $\mathrm{[Fe/H]}\lesssim -2$ are likely overestimated and are not recommended for use.

    \item The distribution of stellar parameters measured by the RVS pipeline shows a deficit of $\log g <2$ stars, likely caused by $\log g$ bias for stars at the top of the red giant branch.
    \item As in the previous version of the pipeline, since RVS pipeline does not have white dwarf models, the spectra of WDs are typically fitted by RVS as $\log g \sim 6$ dwarf stars with high $v\sin i$ to compensate for pressure broadening in WD. 
    \item Some hot $T_\mathrm{eff}>10^4$\,K stars with well-fitted spectra by RVS may have {\tt RVS\_WARN} that is non-zero. Specifically the 4-th bit of the bitmask is on, i.e., corresponding to {\tt RVS\_WARN}=8. This is because metallicity is challenging to measure for hot stars, and the pipeline hit the [Fe/H] grid edge at [Fe/H]=$-4$ triggering the 4-th bit of the {\tt RVS\_WARN} mask. That does not mean that the velocity is incorrect, but other stellar parameters are potentially suspicious.
    \item A small number of sources (fraction of a percent) in the RVS catalogue have non-finite (NaN) uncertainties in $\log g$, $T_\mathrm{eff}$, ${\rm [\alpha/Fe]}$ due to a non-positive-definite estimated Hessian matrix of the likelihood at the best-fit parameters. Most of these (>90\%) are not stars. A handful that are stars have low signal-to-noise of around four. We suggest to exclude those. 
    \item The $v \sin i$ measurements from RVS are not reliable as the majority of stars have $v \sin i$ close to zero, and only very large values are an indication of large spectral line broadening.
    \item The SP pipeline measurements based on the PHOENIX grid should not be used (the problematic measurements can be identified using {\tt BESTGRID}=`s\_desi1'). 
    \item The only individual element abundances that are recommended from the SP catalogues are Mg, Ca, and Fe.
\end{itemize}

\section{Validation of $\alpha$ and individual element abundances}
\label{sec:appendix_alpha_elt_abundances}
\begin{figure}
    \centering
    \includegraphics{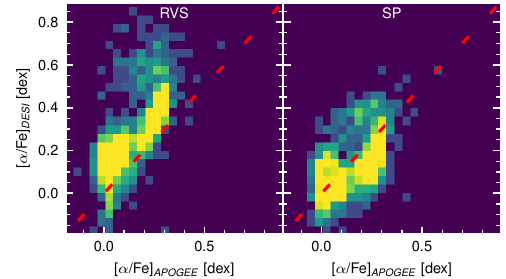}
    \caption{The comparison of $\rm{[\alpha/Fe]}$ measurements from the RVS and SP pipelines with high-resolution literature values from APOGEE.}
    \label{fig:alpha_compar}
\end{figure}

\begin{figure}
    \centering
    \includegraphics{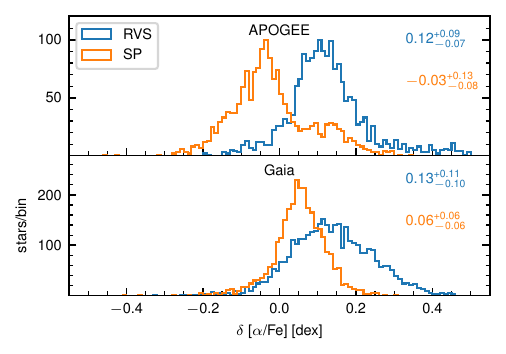}
    \caption{The distribution of differences of DESI $\rm{[\alpha/Fe]}$  measurements with respect to literature values from APOGEE and {\it Gaia}. The top panel shows comparison with APOGEE, the bottom shows {\it Gaia}. Different curves show the measurements from either RVS or SP pipelines.}
    \label{fig:alpha_compar_delta}
\end{figure}

Similar to the discussion of [Fe/H] validation in Section~\ref{sec:feh_compar}, here we provide the comparison of  $\rm{[\alpha/Fe]}$  measured by both SP and RVS pipelines and individual element abundances of Ca and Mg measured by SP with high-resolution literature values. Figures~\ref{fig:alpha_compar} and \ref{fig:alpha_compar_delta} are analogous to Figures~\ref{fig:feh_compar1} and \ref{fig:feh_compar2} but for $\rm{[\alpha/Fe]}$ measurements from both RVS and SP pipelines compared to APOGEE and {\it Gaia} values. We use the same set of selection cuts as for the [Fe/H] validation described in Section~\ref{sec:feh_compar}.

This comparison shows that the RVS $\rm{[\alpha/Fe]}$ values have a systematic offset of $\sim 0.12$\,dex, with a smaller offset of $-0.03$\,dex for the SP pipeline measurements. The scatter of the residuals is $\sim 0.1$\,dex for both pipelines when comparing to APOGEE data. For the  {\it Gaia} {\tt gspspec} data, SP $\rm{[\alpha/Fe]}$  shows much better agreement  with a scatter of 0.06\,dex.  The distribution of residuals between DESI and APOGEE measurements shows signs of bimodality and extended tails, indicative that DESI  $\rm{[\alpha/Fe]}$ systematic errors are likely correlated with other stellar properties like [Fe/H], $\log g$  or $T_{\rm eff}$. We also remark that the validation sample is quite metal-rich with median ${\rm [Fe/H]}\sim -0.5$ for APOGEE and {\it Gaia}, hence it is difficult to assess how representative the validation of $\rm{[\alpha/Fe]}$ is for more metal-poor stars.


\begin{figure} 
    \centering
    \includegraphics{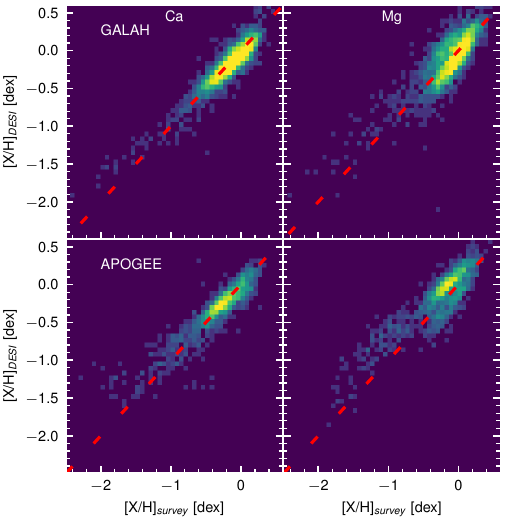}
    \caption{The comparison of Ca and Mg element abundances measured by the DESI SP pipeline with APOGEE and GALAH high-resolution measurements.}
    \label{fig:elt_compar}
\end{figure}

\begin{figure}
    \centering
    \includegraphics{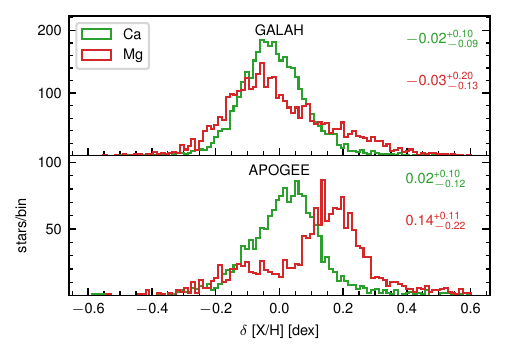}
    \caption{The distribution of differences of [Ca/H] and [Mg/H] elemental abundances measured by the SP pipeline and high-resolution measurements. The top panel shows differences with respect to GALAH, the bottom differences with respect to APOGEE. Different curves show either [Ca/H] (green) or [Mg/H] (red).}
    \label{fig:elt_compar_delta}
\end{figure}

Figures~\ref{fig:elt_compar} and \ref{fig:elt_compar_delta} show the comparison of [Mg/H] and [Ca/H] measured by the SP pipeline with APOGEE and GALAH measurements.
 These abundances are stored in the {\tt ELEM} column of the SP measurement table. As the {\tt ELEM} column stores the abundances of multiple elements as an array, Ca is the 2nd and Mg is 4th elements in the array.
The comparison shows that Calcium measurements (green curves in Figure~\ref{fig:elt_compar_delta}) are the most robust, with median offset with respect to GALAH and APOGEE of $\lesssim 0.02$\,dex and typical scatter with respect to high-resolution measurement of 0.1 dex. The Mg measurements show signs of bimodality of the residuals (see red curves in Figure~\ref{fig:elt_compar_delta} and right columns in Figure~\ref{fig:elt_compar}) and also show significant systematic offset of $0.13$\,dex with respect to APOGEE measurements and scatter of up to $\sim 0.2$\,dex.


\section{Radial velocity systematics for the backup program in DR1}
\label{sec:appendix_backup}

\begin{figure*}
    \centering
    \includegraphics{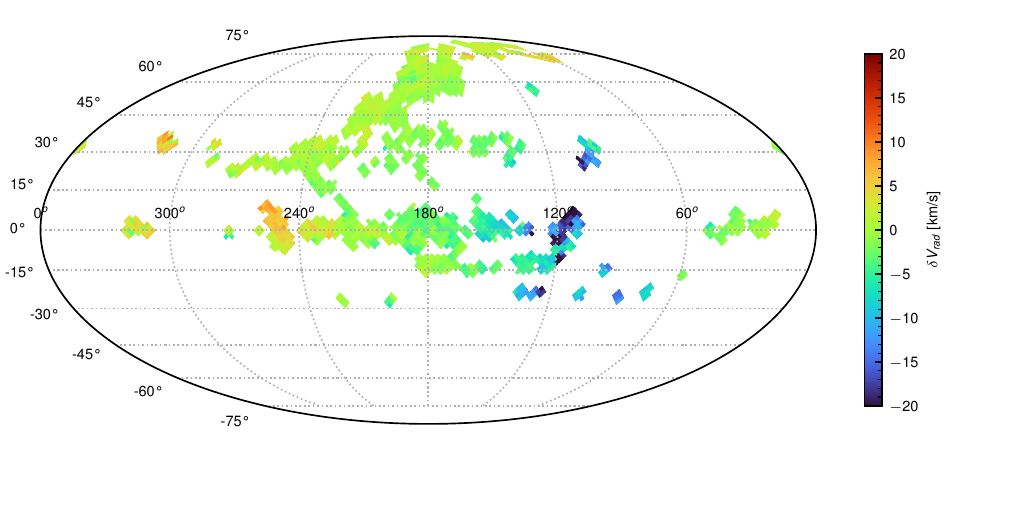}
    \caption{Median offset of DESI radial velocities in the backup program with respect to {\it Gaia} RVS shown as a function of position on the sky in Equatorial coordinates.}
    \label{fig:backup_systematics}
\end{figure*}

As mentioned in the main text the backup program data in DR1 suffers from systematics in radial velocities that are caused by the DESI wavelength calibration issues in poor observing conditions. Because they are specific to individual DESI tiles, the velocity biases have a systematic pattern across the sky.
Figure~\ref{fig:backup_systematics} shows the map of the median offset of DESI velocities with respect to {\it Gaia} DR3 RVS radial velocity measurements. The spatial pattern seen in the Figure is the result of the correlation of the position on the sky and the time when the tile was observed.
Given that the systematics are correlated with the  DESI tile number (i.e., each DESI pointing can approximately be assumed to have similar systematic), we provide a simple correction for users of the VAC. The correction table is based on computing the median deviation of DESI velocities with respect to {\it Gaia} RVS within each DESI tile. We then assign each source in the backup program the median offset of the tile the source is in. Top few rows of the velocity correction table are given in Table~\ref{tab:backup_offsets}. And the full table is published on Zenodo\url{https://doi.org/10.5281/zenodo.15469272}.
The offset needs to be subtracted:
\begin{align}
    V_\mathrm{corrected} = V_\mathrm{RVS} - \mathrm{VRAD\_OFFSET}    
    \label{eq:backup_offset}
\end{align}
16\% of sources have an offset below $-4.7 $\kms and 16\% of  sources have offset larger than $1.7$\kms.

\begin{table}
    \centering
    \begin{tabular}{c|c}
    \hline
    TARGETID & VRAD\_OFFSET\\
     & km\, s$^{-1} $ \\
     \hline
    2305843036808031251 & -3.062\\
    2305843036808030356 & -3.062 \\
... & ... \\
\hline
    \end{tabular}
    \caption{The radial velocity offset for each target in the backup program. The offset needs to be subtracted from radial velocities (see Equation~\ref{eq:backup_offset})}
    \label{tab:backup_offsets}
\end{table}

\section{Metallicity distributions of substructures in DESI}
\label{sec:appendix_fehs}
Here we provide the measurement of mean [Fe/H] and [Fe/H] intrinsic spread in a variety of systems observed by DESI described in Section~\ref{sec:discussion} and Table~\ref{tab:members}.
For each system, we use potential members selected according to the process described in Section \ref{sec:discussion}, and fit the distribution of [Fe/H] after applying the calibration from Section~\ref{sec:feh_calib} by a mixture model for the metallicity $z={\rm [Fe/H]}$

\begin{align}
    P(z|f_{bg},z_{bg},\sigma_{bg},z_{obj},\sigma_{obj}) = f_{bg} \cdot {\mathcal N}(z|z_{bg},\sigma_{bg})+\\
    (1-f_{bg})\cdot {\mathcal N}(z|z_{obj},\sigma_{obj})
\end{align}
with the following priors $f_{bg} \sim U(0,0.3)$ for the background fraction, $\sigma_{bg}\sim U(0.7,1.3)$ for the [Fe/H] scatter of the background stars, $\sigma_{obj}\sim U(0.02,0.7)$ for the [Fe/H] scatter of the substructure, 
$z_{bg}\sim U(-2,0)$ for the mean [Fe/H] of the background and  $z_{obj}\sim U(-4,0.5)$ for the mean [Fe/H] of the substructure.
The posteriors were sampled using {\tt dynesty} nested sampling code \citep{dynesty1,dynesty2}. When we evaluate the likelihood we take into account the individual [Fe/H] uncertainties reported by the pipelines. Table~\ref{tab:members_full} gives the measurement for mean metallicities and intrinsic scatter of [Fe/H] for both SP and RVS pipelines. In a handful of cases there were fewer than 10 stars satisfying the quality cuts; in such cases we do not provide the measurement. We also inspected the fitted [Fe/H] distributions and in the cases of LMS-1 and Leiptr, the metallicity distributions were very broad and lacked a clear peak, so we do not provide the measurement for these systems as a precaution. 

The mean metallicities for the globular clusters observed by DESI are in good (0.1--0.2 dex) agreement with the mean metallicities reported in the literature by \citet{harris1996}. The mean metallicities for the dwarf galaxies, on the other hand, are systematically $\sim$0.2--0.3 dex lower than those reported by \citet{simon2019}. We believe  \citep[see][]{Ding2025} that the reason for this is that DESI's samples are typically more extended than existing spectroscopic samples and thus are probing metal-poorer outer parts of dwarfs.
As expected, the table shows the abundance spreads that are typically lower for GCs and higher for dwarfs \citep{willman2012}. There are, however, few cases where the measured [Fe/H] scatter for a GC is unrealistically high ($>$0.2 dex, i.e. M 3). Also for a few systems (e.g. M 92 and Sextans) the RVS abundances show higher measured intrinsic scatter compared to SP. These discrepancies likely indicate either additional systematics in the calibrated [Fe/H] abundances or an imperfect mixture model. We leave the investigation of that for future papers. In general, we find good agreement between mean [Fe/H] measured by the RVS and SP pipelines, with the worst disagreement for NGC 5634 of 0.21 dex.

\begin{table*}
\centering
\caption{The measured mean [Fe/H] and intrinsic scatter of [Fe/H] for different Milky Way substructures and satellites observed  by DESI. The measurements are provided for  RVS and SP pipelines. The measured values are medians of the posterior and include the uncertainties estimated from 16/84th percentiles of the posterior.}
\label{tab:members_full}
\begin{tabular}{lllll|lllll}
\hline
\textbf{Name} & \textbf{[Fe/H]$_\mathrm{RVS}$} & \textbf{$\sigma$[Fe/H]$_\mathrm{RVS}$} & \textbf{[Fe/H]$_\mathrm{SP}$} & \textbf{$\sigma$[Fe/H]$_\mathrm{SP}$} & \textbf{Name} & \textbf{[Fe/H]$_\mathrm{RVS}$} & \textbf{$\sigma$[Fe/H]$_\mathrm{RVS}$} & \textbf{[Fe/H]$_\mathrm{SP}$} & \textbf{$\sigma$[Fe/H]$_\mathrm{SP}$} \\
\hline
Canes Venatici 1 &  &  & $-2.19_{-0.15}^{+0.15}$ & $0.42_{-0.11}^{+0.13}$ & C-25 & $-2.24_{-0.07}^{+0.06}$ & $0.19_{-0.06}^{+0.07}$ & $-2.13_{-0.07}^{+0.08}$ & $0.19_{-0.05}^{+0.07}$ \\
Draco 1 & $-2.10_{-0.04}^{+0.03}$ & $0.31_{-0.04}^{+0.04}$ & $-2.20_{-0.04}^{+0.03}$ & $0.36_{-0.03}^{+0.03}$ & C-10 & $-0.65_{-0.04}^{+0.03}$ & $0.15_{-0.03}^{+0.05}$ & $-0.76_{-0.03}^{+0.04}$ & $0.06_{-0.03}^{+0.06}$ \\
Sextans 1 & $-2.08_{-0.04}^{+0.04}$ & $0.47_{-0.04}^{+0.03}$ & $-2.24_{-0.04}^{+0.03}$ & $0.44_{-0.03}^{+0.03}$ & Gjoll & $-1.50_{-0.05}^{+0.05}$ & $0.15_{-0.05}^{+0.06}$ & $-1.57_{-0.05}^{+0.04}$ & $0.10_{-0.06}^{+0.07}$ \\
Ursa Major 2 &  &  &  &  & C-24 & $-0.53_{-0.03}^{+0.02}$ & $0.12_{-0.03}^{+0.03}$ & $-0.58_{-0.02}^{+0.03}$ & $0.09_{-0.02}^{+0.03}$ \\
NGC 2419 & $-2.09_{-0.07}^{+0.06}$ & $0.19_{-0.05}^{+0.06}$ & $-2.19_{-0.03}^{+0.03}$ & $0.09_{-0.03}^{+0.05}$ & Slidr & $-1.39_{-0.05}^{+0.05}$ & $0.33_{-0.04}^{+0.04}$ & $-1.55_{-0.06}^{+0.07}$ & $0.33_{-0.04}^{+0.05}$ \\
M 53 & $-1.92_{-0.02}^{+0.03}$ & $0.13_{-0.03}^{+0.03}$ & $-2.00_{-0.01}^{+0.02}$ & $0.06_{-0.01}^{+0.02}$ & Ylgr & $-1.68_{-0.09}^{+0.09}$ & $0.30_{-0.09}^{+0.12}$ & $-1.78_{-0.07}^{+0.06}$ & $0.19_{-0.04}^{+0.07}$ \\
NGC 5053 & $-2.46_{-0.04}^{+0.05}$ & $0.26_{-0.03}^{+0.04}$ & $-2.42_{-0.04}^{+0.04}$ & $0.18_{-0.04}^{+0.03}$ & Sylgr & $-2.47_{-0.03}^{+0.03}$ & $0.14_{-0.03}^{+0.03}$ & $-2.55_{-0.02}^{+0.02}$ & $0.07_{-0.02}^{+0.02}$ \\
M 3 & $-1.65_{-0.04}^{+0.03}$ & $0.26_{-0.03}^{+0.03}$ & $-1.70_{-0.03}^{+0.03}$ & $0.15_{-0.02}^{+0.03}$ & New-15 & $-1.96_{-0.03}^{+0.02}$ & $0.05_{-0.02}^{+0.04}$ & $-2.00_{-0.04}^{+0.04}$ & $0.09_{-0.04}^{+0.04}$ \\
NGC 5466 & $-1.93_{-0.04}^{+0.03}$ & $0.10_{-0.04}^{+0.04}$ & $-2.11_{-0.03}^{+0.03}$ & $0.09_{-0.02}^{+0.03}$ & Gaia-7 & $-0.79_{-0.04}^{+0.04}$ & $0.13_{-0.04}^{+0.03}$ & $-0.82_{-0.05}^{+0.04}$ & $0.13_{-0.03}^{+0.04}$ \\
NGC 5634 & $-1.72_{-0.03}^{+0.04}$ & $0.10_{-0.04}^{+0.04}$ & $-1.93_{-0.06}^{+0.06}$ & $0.24_{-0.04}^{+0.06}$ & Fjorm & $-2.19_{-0.03}^{+0.04}$ & $0.11_{-0.04}^{+0.04}$ & $-2.29_{-0.03}^{+0.03}$ & $0.06_{-0.03}^{+0.04}$ \\
M 5 & $-1.31_{-0.01}^{+0.00}$ & $0.10_{-0.00}^{+0.01}$ & $-1.42_{-0.01}^{+0.00}$ & $0.10_{-0.00}^{+0.01}$ & Orphan & $-1.95_{-0.06}^{+0.07}$ & $0.29_{-0.05}^{+0.07}$ & $-2.01_{-0.08}^{+0.08}$ & $0.32_{-0.06}^{+0.08}$ \\
M 13 & $-1.57_{-0.01}^{+0.01}$ & $0.11_{-0.01}^{+0.01}$ & $-1.66_{-0.01}^{+0.01}$ & $0.09_{-0.01}^{+0.02}$ & Gaia-1 & $-1.41_{-0.02}^{+0.03}$ & $0.10_{-0.02}^{+0.03}$ & $-1.47_{-0.03}^{+0.02}$ & $0.08_{-0.02}^{+0.03}$ \\
M 12 & $-1.33_{-0.03}^{+0.03}$ & $0.09_{-0.02}^{+0.04}$ & $-1.41_{-0.04}^{+0.04}$ & $0.11_{-0.03}^{+0.03}$ & LMS-1 &  &  &  &  \\
M 92 & $-2.25_{-0.03}^{+0.03}$ & $0.25_{-0.05}^{+0.03}$ & $-2.22_{-0.01}^{+0.02}$ & $0.07_{-0.02}^{+0.03}$ & GD-1 & $-2.30_{-0.03}^{+0.02}$ & $0.22_{-0.03}^{+0.03}$ & $-2.24_{-0.02}^{+0.02}$ & $0.13_{-0.02}^{+0.02}$ \\
M 15 & $-2.44_{-0.05}^{+0.06}$ & $0.26_{-0.07}^{+0.10}$ & $-2.38_{-0.07}^{+0.06}$ & $0.15_{-0.05}^{+0.06}$ & Gaia-6 & $-1.19_{-0.03}^{+0.05}$ & $0.08_{-0.03}^{+0.09}$ & $-1.20_{-0.07}^{+0.06}$ & $0.21_{-0.10}^{+0.09}$ \\
M 2 & $-1.56_{-0.02}^{+0.03}$ & $0.15_{-0.02}^{+0.03}$ & $-1.72_{-0.03}^{+0.03}$ & $0.16_{-0.03}^{+0.02}$ & New-19 & $-0.65_{-0.04}^{+0.04}$ & $0.11_{-0.03}^{+0.04}$ & $-0.84_{-0.09}^{+0.10}$ & $0.26_{-0.06}^{+0.09}$ \\
Pal 5 & $-1.23_{-0.03}^{+0.03}$ & $0.14_{-0.03}^{+0.03}$ & $-1.41_{-0.05}^{+0.05}$ & $0.23_{-0.03}^{+0.04}$ & Pal 5 & $-1.20_{-0.04}^{+0.03}$ & $0.16_{-0.03}^{+0.04}$ & $-1.24_{-0.05}^{+0.05}$ & $0.19_{-0.04}^{+0.05}$ \\
NGC 2632 &  &  &  &  & Svol & $-1.38_{-0.14}^{+0.14}$ & $0.43_{-0.14}^{+0.11}$ & $-1.47_{-0.17}^{+0.13}$ & $0.50_{-0.10}^{+0.10}$ \\
Melotte 111 &  &  &  &  & M 92 & $-2.37_{-0.09}^{+0.12}$ & $0.33_{-0.10}^{+0.13}$ & $-2.33_{-0.09}^{+0.12}$ & $0.22_{-0.12}^{+0.16}$ \\
NGC 2539 & $-0.08_{-0.02}^{+0.02}$ & $0.10_{-0.02}^{+0.02}$ & $-0.07_{-0.02}^{+0.02}$ & $0.08_{-0.02}^{+0.02}$ & Hrid & $-1.13_{-0.03}^{+0.03}$ & $0.11_{-0.03}^{+0.02}$ & $-1.24_{-0.03}^{+0.04}$ & $0.11_{-0.03}^{+0.05}$ \\
Kwando & $-2.08_{-0.09}^{+0.09}$ & $0.23_{-0.07}^{+0.11}$ & $-2.13_{-0.04}^{+0.05}$ & $0.12_{-0.04}^{+0.06}$ & New-21 & $-0.51_{-0.05}^{+0.04}$ & $0.24_{-0.04}^{+0.05}$ & $-0.52_{-0.05}^{+0.04}$ & $0.18_{-0.03}^{+0.04}$ \\
Leiptr &  &  &  &  & Phlegethon & $-1.63_{-0.03}^{+0.04}$ & $0.16_{-0.04}^{+0.03}$ & $-1.71_{-0.04}^{+0.04}$ & $0.13_{-0.04}^{+0.05}$ \\
\hline
\end{tabular}
\label{tab:mean_substr_feh}
\end{table*}
\end{document}